\documentclass[aps,prd,10pt,preprintnumbers,twocolumn,superscriptaddress,floatfix,nofootinbib,amsmath,amssymb,altaffilletter]{revtex4-1}
\usepackage[colorlinks=true,pdfstartview=FitV,breaklinks=true]{hyperref}

\usepackage{bm}
\usepackage{graphicx}
\usepackage{multirow}
\usepackage{soul}
\usepackage{amsmath}
\usepackage{mathrsfs}
\usepackage{amssymb}
\usepackage{yfonts}
\usepackage{color}
\usepackage{xspace}
\usepackage{url}
\usepackage{verbatim}
\usepackage{mathtools}
\usepackage{upgreek}
\usepackage{units}
\usepackage{siunitx}
\usepackage{amstext}
\usepackage[sc]{mathpazo}
\usepackage{booktabs}
\usepackage{tabulary}
\usepackage{tabularx}
\usepackage{etoolbox}
\usepackage[utf8]{inputenc}
\usepackage[dvipsnames,table]{xcolor}
\newcolumntype{Y}{>{\centering\arraybackslash}X}
\hypersetup{urlcolor=BlueViolet,
	    citecolor=Plum,
	    linkcolor=PineGreen}
\usepackage[official]{eurosym}
\definecolor{lightgray}{rgb}{0.9,0.9,0.9}	    
\definecolor{green}{rgb}{0,0.5,0}
\definecolor{red}{rgb}{1,0,0}
\definecolor{blue}{rgb}{0,0,0.5}

\newcommand{\qrhat}{\hat{\mathbf{q}}_r}
\newcommand{\qnuhat}{\hat{\mathbf{q}}_\nu}
\newcommand{\vlab}{\mathbf{v}_\mathrm{lab}}
\newcommand{\vmin}{v_{\rm min}}

\newcommand{\vesc}{v_\mathrm{esc}}

\newcommand{\kms}{\textrm{ km s}^{-1}}

\newcommand{\dbd}[2]{\ifmmode \frac{\textrm{d}#1}{\textrm{d}#2}\else $\textrm{d}#1/\textrm{d}#2$\fi}
\newcommand{\pbp}[2]{\ifmmode \frac{\partial#1}{\partial#2}\else $\partial#1/\partial#2$\fi}

\newcommand{\ra}[1]{\renewcommand{\arraystretch}{#1}}

\newcommand{\vbf}{\mathbf{v}}

\newcommand{\drm}{\mathrm{d}}

\DeclareMathAlphabet{\mathpzc}{OT1}{pzc}{m}{it}
 \newcommand{\eV}{\text{e\kern-0.15ex V}\xspace}
 \newcommand{\keV}{\text{k\eV}\xspace}
 
 \newcommand{\GeV}{\text{G\eV}\xspace}
 \newcommand{\TeV}{\text{T\kern-0.1ex \eV}\xspace}
 
 \newcommand{\cevns}{CE$\nu$NS\xspace}
  \newcommand{\sigmapSI}{\sigma^{\rm SI}_p}

\newcommand{\Boron}{$^8$B\xspace}

\newcommand{\be}{\begin{equation}}
\newcommand{\ee}{\end{equation}}
\newcommand{\bea}{\begin{eqnarray}}
\newcommand{\eea}{\end{eqnarray}}

\begin{document}
\preprint{CPPC-2020-13}

\title{Can we overcome the neutrino floor at high masses?}

\author{Ciaran A. J. O'Hare}\email{ciaran.ohare@sydney.edu.au}
\affiliation{Sydney Consortium for Particle Physics and Cosmology, University of Sydney, School of Physics, NSW 2006, Australia}
\date{\today}
\smallskip
\begin{abstract}
The neutrino floor is a barrier in the parameter space of weakly interacting massive particles (WIMPs) below which discovery is impeded due to an almost irreducible background of neutrinos. 
Directional gas time projection chambers could discriminate against solar neutrinos, relevant for WIMP masses $\lesssim$10~GeV. 
At higher masses $\gtrsim$100~GeV the floor is set by the background of atmospheric neutrinos. Probing below this part of the floor would require very large target exposures. 
Since gas-based detectors would be prohibitively large at this scale, we instead reevaluate the prospects for liquid noble experiments to probe below the neutrino floor.
We combine all potential methods of subtracting the neutrino background to determine how much of this difficult to reach, but well-motivated, parameter space it is feasible to reach. 
Most notably, we quantify whether a proposed directional signal in xenon and argon experiments called ``columnar recombination'' can help in this task. 
We find that even if the strength of this effect is amplified beyond current experimental results, the quantity of directional information contained in the recombination signal is too low to realistically discriminate against the atmospheric neutrino background.
Instead, benefiting from the refined measurements of neutrino fluxes by experiments such as DUNE and JUNO will be the most practical means to push direct WIMP searches below the neutrino floor. 
For an ultimate global coordination of xenon and argon experiments, we show that the neutrino floor is a surmountable barrier. 
The direct detection of 100 GeV-scale supersymmetric WIMPs may, eventually, be within reach.
\end{abstract}

\maketitle

\section{Introduction}
A rapid succession of increasingly stringent null results from direct dark matter (DM) detection experiments (see Refs.~\cite{Undagoitia:2015gya,Schumann:2019eaa,Battaglieri:2017aum} for reviews) has ruled out swathes of preferred WIMP parameter space~\cite{Arcadi:2017kky}. Even further increases in sensitivity are expected over the next few years, especially as liquid xenon (LXe) and argon (LAr)-based detectors reach the multiton-scale. These colossal detectors will be so large and so sensitive that they are poised to detect coherent neutrino-nucleus scattering (\cevns) for the first time with a natural source of neutrinos. Cosmic, terrestrial and human-made neutrinos form the ultimate background for DM searches on Earth~\cite{Monroe:2007xp,Vergados:2008jp,Strigari:2009bq,Gutlein:2010tq,Billard:2013qya}.

In a typical direct detection experiment the neutrino background looks similar to a WIMP signal, and cannot be shielded. The only discoverable cross sections for WIMP masses which have signals that are mimicked by such a background are those that can provide an excess in events larger than the expected size of potential statistical fluctuations of that background. For the neutrino background, the dominant uncertainty is the systematic uncertainty on the various neutrino flux normalisations, which range from 1\%--50\% depending on the source of neutrino. The cross section below which the WIMP signal is saturated by this uncertainty is labelled the ``neutrino floor''~\cite{Billard:2013qya}. Since 2013 it has been shown underneath all experimental results, often billed as the ultimate limit to conventional direct DM detection. Just like a generic WIMP limit, the shape of the neutrino floor is dependent on nuclear~\cite{Papoulias:2018uzy}, astrophysical~\cite{OHare:2016pjy} and particle model~\cite{Dent:2016iht, Dent:2016wor, Gelmini:2018ogy} inputs for the WIMP signal, but can also be modified~\cite{AristizabalSierra:2017joc,  Gonzalez-Garcia:2018dep}, and even raised by several orders of magnitude~\cite{Boehm:2018sux}, if there are any non-standard neutrino-nucleus interactions.

To circumvent the neutrino floor, an experiment requires some form of discriminating information. Previous work has shown that if the directional dependence of both the WIMP signal and the neutrino background can be measured then this information can help to set limits beyond the neutrino floor~\cite{Grothaus:2014hja,O'Hare:2015mda,Mayet:2016zxu,Franarin:2016ppr,OHare:2017rag}. 
Independently of the neutrino background, directional signals are highly sought-after in direct detection experiments in general because they offer a means to definitively test for the galactic origin of a signal and confirm it to be DM~\cite{Mayet:2016zxu}, as well as to measure the local velocity structure of the Milky Way's DM halo~\cite{Billard:2012qu,Lee:2012pf,O'Hare:2014oxa,Kavanagh:2016xfi,OHare:2018trr,OHare:2019qxc}.
So far however directional detection has only been shown to be experimentally feasible for detectors using low-density targets, the most discussed example being gas time projection chambers (TPCs)~\cite{Battat:2016pap}. While gas targets have various limitations, for probing beyond the neutrino floor a positive cost-balanced trade-off is possible when the focus is shifted towards low energy thresholds, short drift lengths and a modular configuration. Gas TPCs are therefore more appropriate for tackling the low mass shoulder of the neutrino floor due to solar neutrinos. This is the aim of the {\sc Cygnus} project~\cite{Vahsen:2020pzb}. 

If the low mass shoulder of the neutrino floor is surmountable, we are left to ask if future experiments will be able to overcome the floor at the high mass frontier. This is where the final stages of proposed detectors with $\sim$100--1000 ton-year exposures like DARWIN~\cite{Aalbers:2016jon} and Argo~\cite{Sanfilippo:2019amq} are projected to reach. Cross sections below the neutrino floor for 100 GeV--TeV masses still retain substantial motivation from a theory standpoint when one invokes supersymmetry (SUSY) as the theoretical origin of the WIMP. Preferred regions for the nucleon scattering cross sections of the lightest neutralinos are frequently found below the neutrino floor for 100 GeV--TeV masses. See e.g.~Refs.~\cite{Athron:2017qdc,Kobakhidze:2018vuy,Roszkowski:2014iqa,Hisano:2011cs,Abdullah:2016avr} for just a few recent examples. Even in the absence of a UV complete DM-producing theory, candidates below the neutrino floor can appear in simplified models and effective theories as well~\cite{Drees:2019qzi,Balkin:2018tma}. For instance, small cross sections are expected naturally if DM-nucleus cross sections are momentum suppressed, as they would be for particle models with a pseudoscalar mediator~\cite{Arina:2019tib,Arcadi:2017wqi}. Even WIMP-like particles with alternative non-thermal production mechanisms can predict low cross sections and high masses at and below the neutrino floor, e.g.~the recently proposed filtered DM~\cite{Baker:2019ndr,Chway:2019kft}.

Experiments utilising gas targets will never be competitive with liquid noble experiments in accessing such low cross sections. The $\sim$100 ton-year target masses needed to reach spin independent WIMP-nucleon cross sections below $10^{-48}$~cm$^2$ would require TPC volumes in excess of 100,000 m$^3$~\cite{Vahsen:2020pzb}, or roughly twice the size of the internal vessel volume proposed for DUNE. While it is always possible to raise the operating pressure to reduce the required volume, this comes at the cost of increased diffusion and consequently poor track reconstruction. Trading off a higher operating pressure with a reduction in drift length is also undesirable because it would lead to extremely large and costly readout planes.

The natural question to ask is then, can directional detection be done in high density targets, without direct track reconstruction? The difficulty is that keV-scale recoils are just too short, e.g.~$\mathcal{O}(10\,{\rm nm})$ in LXe or LAr. Any directional information will have to be extracted indirectly when not using gas. One suggestion in the context of liquid noble gas detectors is to exploit an effect known as columnar recombination~\cite{Nygren:2013nda}. This is a process that could generate an asymmetry between the ionisation and scintillation yields of recoils that point parallel or perpendicular to an applied electric field. Knowledge of this effect dates back over a century~\cite{Jaffe:1913} and has been observed in both xenon and argon, albeit at higher energies than are relevant for a DM search~\cite{Nakajima:2015dva,Nakamura:2018xvy,MuAaoz:2014uxa}. To  generate a usable level of directionality, recoil tracks must be long relative to the typical length scale for electrons and ions to recombine. For xenon this implies that columnar recombination is probably unobservable at keV energies in liquid (instead one would need a high-pressure gas mode at $\sim$10 bars~\cite{Nygren:2013nda}), but there is still some hope for argon. Ongoing investigation by ReD~\cite{Cadeddu:2017ebu,Cadeddu:2016mac}---as part of DarkSide, but continuing the work of SCENE~\cite{Cao:2014gns}---aims to determine the feasibility of using the effect in a LAr DM search. 

We aim here to determine if columnar recombination will help future multiton scale experiments to probe beyond the neutrino floor. In the process we will present a more detailed calculation of the atmospheric neutrino background for direct DM experiments than considered previously, focusing on its angular dependence. We will also not only incorporate directionality, but all possible discriminants that could be used in future LXe or LAr experiments to overcome the neutrino floor.


The paper is structured as follows: firstly, in Sec.~\ref{sec:theory} we review the neutrino background to direct DM searches. Then in Sec.~\ref{sec:neutrinofloor} we briefly outline how to calculate discovery limits, and discuss various subtleties around how the neutrino background really impacts the discovery of DM. In Sec.~\ref{sec:directionality} we introduce the concept of directionality and develop a simple model for columnar recombination as a possible example in liquid noble gas experiments. In Sec.~\ref{sec:results} we show our final results and in Sec.~\ref{sec:summary} we summarise and conclude. The code used to produce our results is available at \url{https://github.com/cajohare/AtmNuFloor}

\section{The neutrino background}\label{sec:theory}

\subsection{Coherent neutrino-nucleus scattering}
The \cevns background in direct DM detection experiments produces \keV-scale nuclear recoils with a spectrum similar to certain WIMP masses. This Standard Model process was only recently observed for the first time by COHERENT~\cite{Akimov:2017ade, Akimov:2020pdx, Akimov:2020czh}. \cevns proceeds via a neutral current and has a coherence effect at low momentum transfer that approximately scales with the number of neutrons squared~\cite{Freedman:1973yd,Freedman:1977, Drukier:1983gj}. At higher recoil energies, generally above a few tens of \keV, the loss of coherence is described by the nuclear form factor $F(E_r)$, for which we use the standard Helm ansatz~\cite{Lewin:1995rx}--an excellent approximation at these still relatively low energies~\cite{Vietze:2014vsa}. 

The differential \cevns cross section as a function of the nuclear recoil energy ($E_r$) and neutrino energy ($E_\nu$) is given by~\cite{Freedman:1973yd,Freedman:1977,Drukier:1983gj}
\begin{equation}\label{eq:CEvNS}
 \dbd{\sigma}{E_r}(E_r,E_\nu) = \frac{G_F^2}{4 \pi} Q^2_W m_N \left(1-\frac{m_N E_r}{2 E_\nu^2} \right) F^2(E_r) \, ,
\end{equation}
where $Q_W = A-Z - (1-4\sin^2\theta_W) Z$ is the weak hypercharge of a nucleus with mass number $A$ and atomic number $Z$, $G_F$ is the Fermi coupling constant, $\sin^2{\theta_W} = 0.2312$ is the weak mixing angle, and $m_N$ is the target nucleus mass. 

The differential cross section as a function of the direction of the recoiling nucleus, $\Omega_r$, can be obtained by first noting that the scattering has azimuthal symmetry about the incoming neutrino direction, i.e.~$\drm\Omega_\nu~=~2\pi \,\drm\cos\beta$. The kinematic expression for the angle, $\beta \in [0,\pi/2]$, between the neutrino direction, $\hat{{\bf q}}_\nu$, and the recoil direction, $\hat{{\bf q}}_r$ is~\cite{Vogel:1989iv},
\begin{equation}\label{eq:kinematics}
 \cos{\beta} = \hat{{\bf q}}_r \cdot \hat{{\bf q}}_\nu = \frac{E_\nu + m_N}{E_\nu}\sqrt{\frac{E_r}{2 m_N}} \, .
\end{equation}
We impose this relation with a delta function to get,
\begin{equation}\label{eq:doublecrosssection}
  \frac{\drm^2 \sigma}{\drm E_r \drm \Omega_r} = \dbd{ \sigma}{E_r} \, \frac{1}{2 \pi}\, \delta\left(\cos\beta - \frac{E_\nu + m_N}{E_\nu} \sqrt{\frac{E_r}{2 m_N}}\right) \,.
\end{equation}
The maximum recoil energy, $E_r^{\rm max}$ corresponds to $\beta = 0$,
\begin{equation}\label{eq:Emax}
E_r^{\rm max} = \frac{2 m_N E_\nu^2}{(E_\nu + m_N)^2} \approx \frac{2E_\nu^2}{m_N + 2E_\nu} \,.
\end{equation}

Since we are interested in directionally sensitive experiments we write down the \cevns event rate per unit detector mass, as a function of the recoil energy, direction and time. This is given by the convolution of the cross section and the neutrino flux $\Phi$ which may be time dependent,
\begin{equation}\label{eq:nu_directionalrate}
  \frac{\drm^2 R_\nu(t)}{\drm E_r \drm\Omega_r} =  \frac{1}{m_N} \int_{E_\nu^{\rm min}} \frac{\drm^2 \sigma}{\drm E_r \drm \Omega_r}\frac{\drm^2 \Phi(t)}{\drm E_\nu \drm\Omega_\nu} \drm E_\nu \drm\Omega_\nu \, ,
\end{equation}
where $E_\nu^{\rm min} = \sqrt{m_NE_r/2}$ is the minimum neutrino energy that can produce a nuclear recoil with energy $E_r$.

\begin{table*}[t]\centering
\ra{1.3}
\begin{tabularx}{0.8\textwidth}{Xl|Y|YY|YY|lYr}
\hline\hline
$\nu$ \bf{type} &  & $E_{\nu}^{\rm{rms}}$ & $E_{\rm Xe}^{\rm med}$ & $E_{\rm Ar}^{\rm med}$ & $E_{\rm Xe}^{\rm max}$ & $E_{\rm Ar}^{\rm max}$ & $\Phi(1\pm \delta\Phi/\Phi)$ &  $\times 10^n$ & Ref.\\
 & & [MeV] & [keV] & [keV] & [keV] & [keV] & \multicolumn{3}{c}{[cm$^{-2}$ s$^{-1}$]}  \\
\hline
\multirow{8}{*}{\bf Solar} &$pp$ & $ 0.280 $ & $ 0.001 $ & $ 0.002 $ & $ 0.003 $ & $ 0.010 $ &  $5.98\left(1\pm 0.006 \right)$& $10^{10}$&\cite{Vinyoles:2016djt} \\
&$pep$ & $ 1.440 $ & $ 0.011 $ & $ 0.034 $ & $ 0.035 $ & $ 0.114 $ &$1.44\left(1\pm 0.01\right)$&$10^8$ & \cite{Vinyoles:2016djt} \\
&$hep$ &  $ 10.29 $ & $ 0.604 $ & $ 2.030 $& $ 5.859 $ & $ 19.367 $ & $7.98\left(1\pm 0.30 \right)$ & $10^3$ &\cite{Vinyoles:2016djt} \\ 
&$^{7}\mathrm{Be}$ & $ 0.384 $ & $ 0.001 $ & $ 0.003 $ & $ 0.002 $ & $ 0.008 $&  $4.93\left(1\pm 0.06 \right)$&$10^8$ & \cite{Vinyoles:2016djt} \\
&$^{7}\mathrm{Be}$ & $ 0.861 $ & $ 0.004 $ & $ 0.013 $& $ 0.012 $ & $ 0.041 $ &  $4.50\left(1\pm 0.06 \right)$&$10^9$ & \cite{Vinyoles:2016djt} \\
&$^{8}\mathrm{B}$ & $ 7.259 $ & $ 0.314 $ & $ 1.046 $& $ 4.443 $ & $ 14.687 $ &$5.16\left(1\pm 0.02\right)$& $10^6$ &\cite{Bergstrom:2016cbh}   \\

&$^{13}\mathrm{N}$ & $ 0.749 $ & $ 0.004 $ & $ 0.017 $& $ 0.024 $ & $ 0.078 $ & $2.78\left(1\pm 0.15\right)$& $10^8$ & \cite{Vinyoles:2016djt}\\
&$^{15}\mathrm{O}$ & $ 1.058 $ & $ 0.008 $ & $ 0.023 $& $ 0.050 $ & $ 0.164 $ & $2.05\left(1\pm 0.17\right)$&$10^8$ &\cite{Vinyoles:2016djt} \\
&$^{17}\mathrm{F}$ & $ 0.801 $ & $ 0.005 $ & $ 0.014 $& $ 0.050 $ & $ 0.166 $ & $5.29\left(1\pm 0.20 \right)$ & $10^6$ &\cite{Vinyoles:2016djt}\\
\hline
\multirow{3}{*}{\bf Geo.} & U & $ 1.051 $ & $ 0.011 $ & $ 0.032 $& $ 0.343 $ & $ 1.135 $ &  $4.34(1\pm0.20)$ &$10^6$ & \\
& Th & $ 0.933 $ & $ 0.010 $ & $ 0.030 $ & $ 0.090 $ & $ 0.299 $&  $4.23(1\pm0.25)$ &  $10^6$ &\cite{Huang:2013} \\
& K & $ 0.801 $ & $ 0.005 $ & $ 0.014 $ & $ 0.031 $ & $ 0.101 $&  $2.05(1\pm0.17)$ & $10^7$ & \\
\hline
\multicolumn{2}{l|}{{\bf Reactor}} & $ 0.817 $ & $ 0.035 $ & $ 0.107 $& $ 2.170 $ & $ 7.173 $ &  $3.06(1\pm0.08)$ &$10^6$ & \cite{Baldoncini:2014vda} \\
\hline
{\bf DSNB}&  & $ 8.781 $ & $ 0.788 $ & $ 2.844 $& $ 138.240 $ & $ 455.660 $ & $ 8.57(1\pm 0.50)$ & $10^1$ &\cite{Beacom:2010kk}\\
\hline
\multicolumn{2}{l|}{{\bf Atmospheric}\quad \,}  & $ 477.9 $ & $ 10.27 $ & $ 63.60 $ & $ >1000 $ & $ >1000 $& $1.07(1\pm 0.25)$ & $10^1$ & \cite{Honda:2011nf}\\
\hline \hline
\end{tabularx}
\caption{All relevant neutrino fluxes for direct DM searches. We write flux normalisations and uncertainties as $\Phi(1~\pm~\frac{\delta \Phi}{\Phi})$ in units of $10^n$~cm$^{-2}$~s$^{-1}$. We also display the root-mean-square neutrino energy, $E_{\nu}^{\rm{rms}}$, and the median of the full \cevns recoil spectra for a xenon and argon target: $E_{\rm Xe}^{\rm{med}}$ and $E_{\rm Ar}^{\rm{med}}$ (i.e. the energy above which 50\% of nuclear recoils scatter). We also display the maximum recoil energies generated by each neutrino flux, calculated using Eq.\eqref{eq:Emax}: $E_{\rm Xe}^{\rm{max}}$ and $E_{\rm Ar}^{\rm{max}}$. Reactor and geoneutrinos are calculated at the location of LNGS.\label{tab:neutrinos}}
\end{table*}
In Table~\ref{tab:neutrinos}, we list the total time-averaged flux, $\Phi$, and its systematic uncertainty, $\delta \Phi$, for each source of neutrino relevant for direct detection experiments: solar, atmospheric, geological, reactor and the diffuse supernova neutrino background (DSNB). See Ref.~\cite{Vitagliano:2019yzm} for a recent review of the spectra of each of these components. We also calculate the median and maximum recoil energies of xenon and argon nuclei. Recoil spectra are exponentially falling and for energies higher than $\sim$200~keV in xenon and $\sim$1000~keV in argon, the event rate is highly suppressed by nuclear form factors $F^2(E_r)\lesssim10^{-3}$. For reactor, geoneutrino, and atmospheric fluxes we assume the experiment is at the Laboratori Nazionali del Gran Sasso (LNGS) where DarkSide-20k will be located. We will now discuss in more detail the components of this background most relevant in this study.

\subsection{Atmospheric neutrinos}
\begin{figure}
\begin{center}
\includegraphics[trim = 0mm 0 0mm 0mm, clip, width=0.49\textwidth]{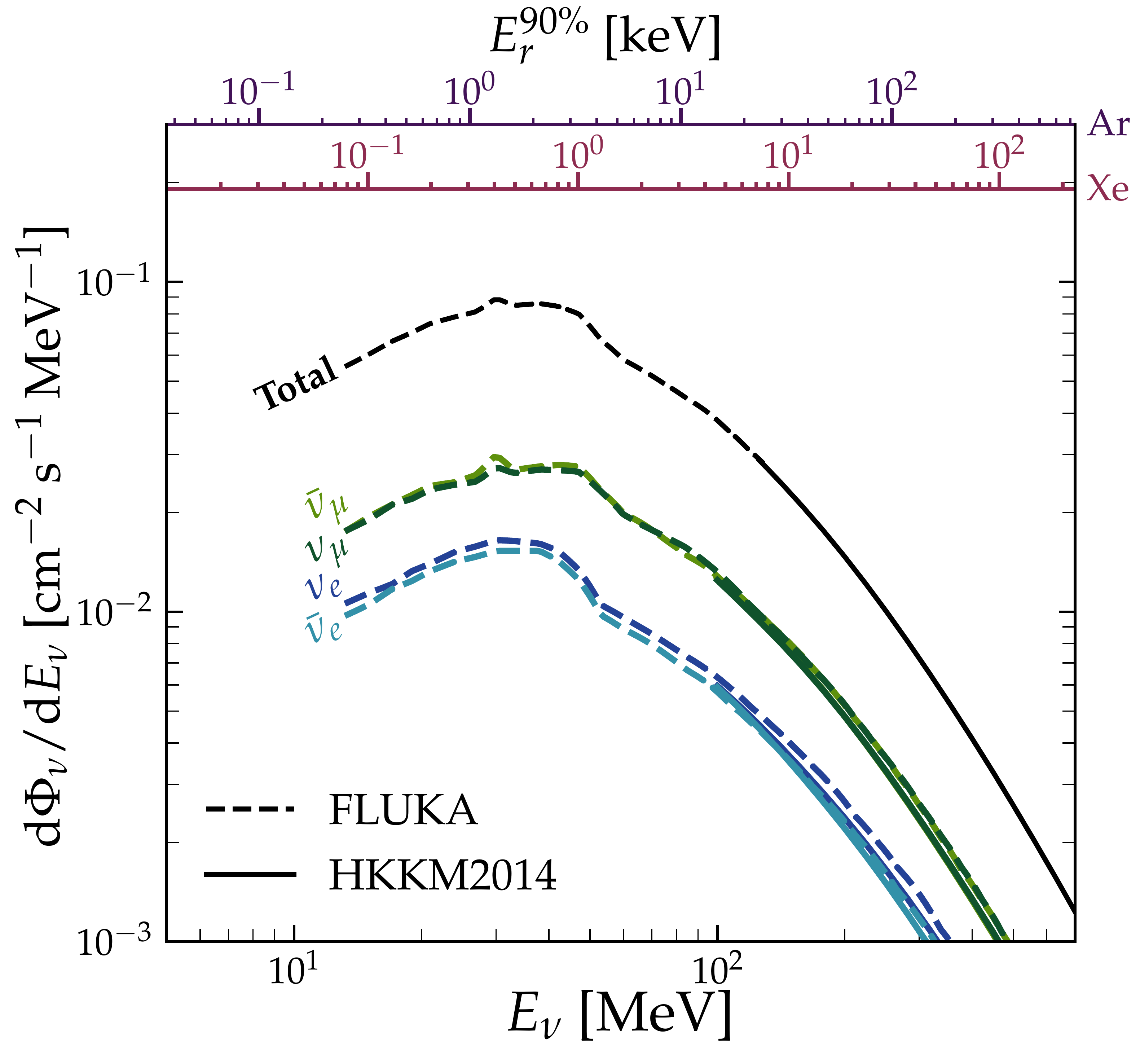}
\caption{The low energy atmospheric neutrino flux. The spectra from 13 MeV--1 GeV are taken from the FLUKA simulation~\cite{Battistoni:2005pd},  (dashed lines), whereas the higher energy components (for which we also have the angular distributions) are from the HKKM2014 simulation~\cite{Honda:2015fha} (solid lines). We show each individual neutrino species ($\nu_e$ and $\bar{\nu}_e$ in blue, and $\nu_\mu$ and $\bar{\nu}_{\mu}$ in green) as well as the total spectrum interpolated between the two simulations (black). On the upper horizontal axis we show $E^{\rm 90\%}_r$ for xenon and argon, which we define to be the energy above which lie 90\% of \cevns recoils for the corresponding incoming neutrino energy.} 
\label{fig:AtmFlux}
\end{center}
\end{figure}

\begin{figure*}
\begin{center}
\includegraphics[trim = 0mm 0 0mm 0mm, clip, width=0.99\textwidth]{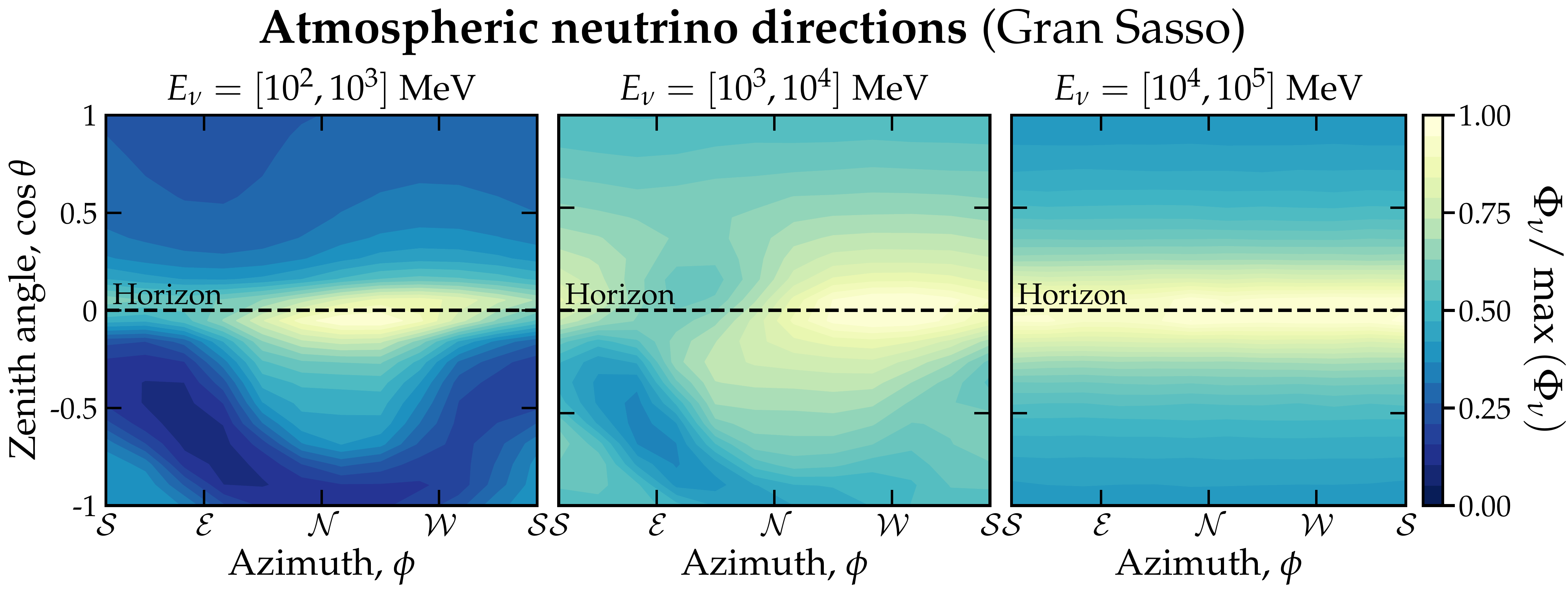}
\caption{Angular distributions of incoming atmospheric neutrino directions integrated over three bins of neutrino energy. We sum over the four species of neutrino. The distributions of arrival angles are given in terms of zenith angle $\cos{\theta}$ and azimuthal angle $\phi$. The angular distributions are from the HKKM2014 simulation~\cite{Honda:2015fha}. We show the full result of this simulation over a wide range of energies to show the general trend in shape of the angular distributions, however only the first panel contributes substantially to the rate of xenon or argon recoils. The distribution is mapped linearly to the colourscale shown to the right of the three panels. Yellow corresponds to the maximum value of $\Phi_\nu$ (across all energies) and dark blue corresponds to $\Phi_\nu = 0$. The colourscale is shared between the three panels.} 
\label{fig:AtmFlux_angle}
\end{center}
\end{figure*} 
 
\begin{figure}
\begin{center}
\includegraphics[trim = 0mm 0 0mm 0mm, clip, width=0.49\textwidth]{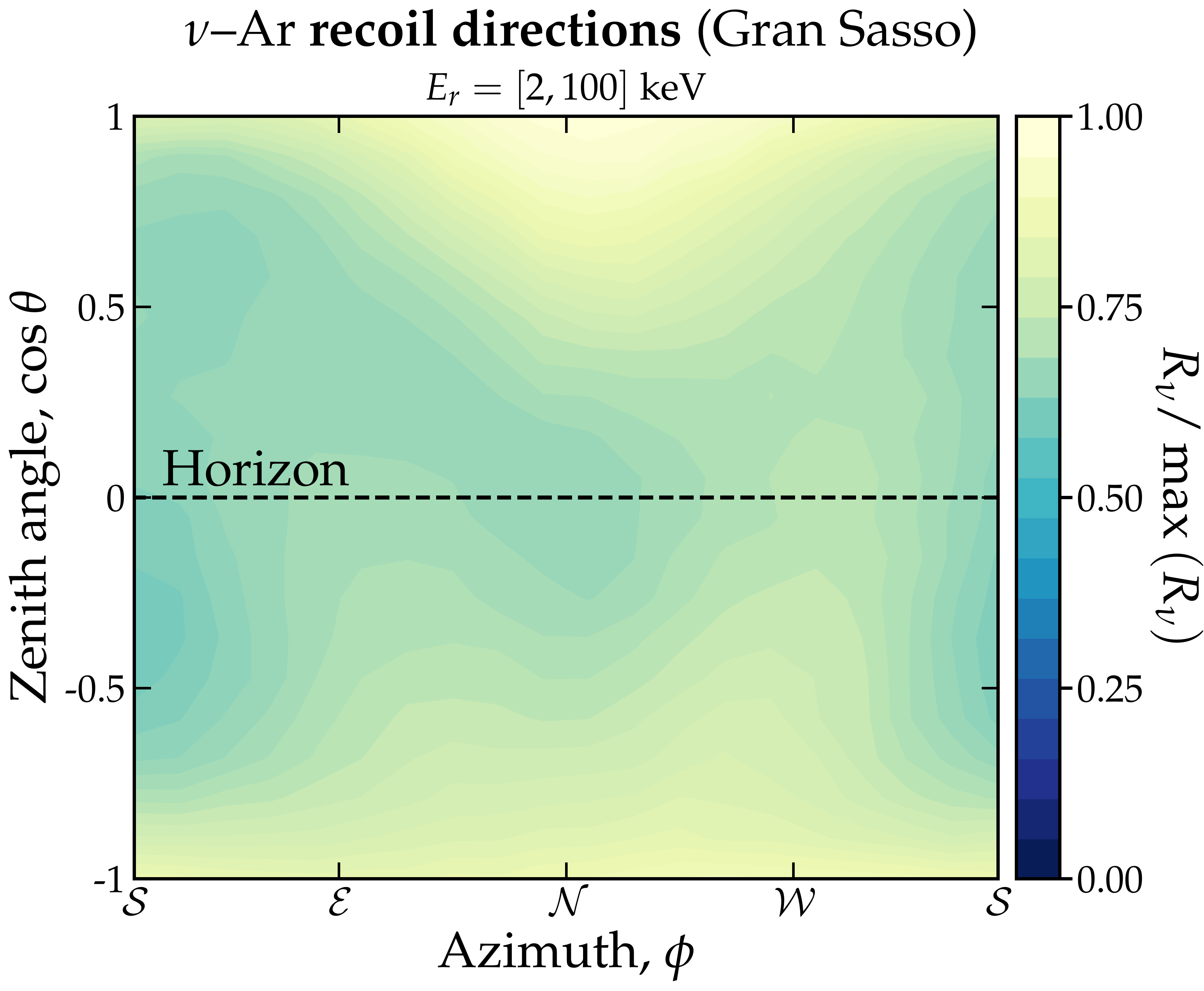}
\caption{Angular distributions of \cevns recoils for an argon target integrated between energies of 2 and 100 keV. As in the previous figure, the angular distributions correspond to the apparent arrival angle of the recoil vector. The recoil distributions are generated from a Monte Carlo simulation of \cevns using incoming neutrino directions drawn from the flux distributions shown previously. The distribution is mapped linearly onto the colourscale shown to the right of the panel, where dark blue corresponds to the maximum value of $R_\nu$ and yellow corresponds to $R_\nu = 0$. Note that the minimum of this distribution is only around 60\% of the maximum value, hence the low contrast compared with the previous figure.} 
\label{fig:AtmFlux_recoilangle}
\end{center}
\end{figure}

Electron and muon neutrinos and antineutrinos are produced in interactions between cosmic rays and particles in the Earth's atmosphere with energies $\gtrsim$10 MeV. Since atmospheric neutrinos are the only source of neutrino with energies from $\sim$50 MeV up to well above a TeV, they alone are responsible for the neutrino floor to WIMP masses $\gtrsim$30 GeV across most targets. For xenon in particular, a 100 GeV WIMP has a recoil spectrum that looks remarkably like the \cevns spectrum from the sub-100 MeV tail of the atmospheric flux~\cite{Strigari:2009bq,Billard:2013qya,Ruppin:2014bra}.

While the energy spectrum of atmospheric neutrinos from 10--100 MeV is well-understood---being simply the spectrum from muon and pion decay---the flux is sensitive to the geomagnetic field and is therefore much more difficult to predict. A well-known flux calculation was made using FLUKA in 2005~\cite{Battistoni:2005pd}, but has uncertainties of up to 25\%. Atmospheric neutrinos with energies of 1 GeV and above are much better understood~\cite{Honda:2011nf,Honda:2015fha}, and have been observed continuously in neutrino experiments since the 1960s. Improved measurements are anticipated in future experiments such as DUNE~\cite{Abi:2020evt,Kelly:2019itm}, Hyper-Kamiokande~\cite{Abe:2018uyc,Li:2017zix}, JUNO~\cite{An:2015jdp,Djurcic:2015vqa,Settanta:2019ecp} and the Jinping Neutrino Experiment~\cite{JinpingNeutrinoExperimentgroup:2016nol}. The sub-100 MeV tail, in particular, is a key limiting background for the highly sought-after, but highly challenging, measurement of the DSNB~\cite{Cocco:2004ac}. LAr TPCs are highly sensitivity to $\nu_e$ at low energies, so measurements of this component of the atmospheric flux should be achievable in DUNE~\cite{Cocco:2004ac,Kelly:2019itm,Capozzi:2018dat}--if potential major sources of background like the muon spallation of argon are well-understood~\cite{Zhu:2018rwc}.

Two predictions for the atmospheric neutrino flux below 10 GeV are shown in Fig.~\ref{fig:AtmFlux}. The lower energy component is the aforementioned FLUKA result~\cite{Battistoni:2005pd}. The higher energy component is from the more recent calculation of Honda et al.~\cite{Honda:2015fha} (HKKM2014), which made use of the updated atmospheric model NRLMSISE--00. Here we adopt their calculation of the time-averaged flux at LNGS. The interpolated all-flavour spectrum (black solid and dashed line) is our input for later calculations. 

The directionality of the atmospheric neutrino flux is dependent on the geomagnetic rigidity cutoff, which varies with position. Cosmic rays with rigidities below the cutoff at a given position will have been deflected. In contrast, those above the cutoff have enough momentum to overcome magnetic deflection and arrive at that position. For neutrinos from cosmic rays more energetic than the geomagnetic rigidity cutoff, the flux peaks along the horizon, $\cos\theta \approx 0$, and is essentially symmetric for $\pm \cos{\theta}$ and independent of azimuthal angle, $\phi$. The flux of neutrinos from cosmic rays below the cutoff on the other hand is asymmetric in both $\cos{\theta}$ and $\phi$ due to the complicated structure of the geomagnetic field. These phenomena are demonstrated in Fig.~\ref{fig:AtmFlux_angle}. We show the angular distribution of the atmospheric neutrino flux from the HKKM2014 simulation~\cite{Honda:2015fha}, integrating over three bins in neutrino energy. We choose a lab-centered coordinate system described by zenith and azimuthal angles, $(\cos{\theta},\phi)$. For the azimuthal angle we display the cardinal direction to avoid potential ambiguity. The angles in Fig.~\ref{fig:AtmFlux_angle} describe the neutrino's arrival direction, $-\mathbf{q}_\nu$, so that $\cos{\theta} = 1$ corresponds to \emph{downward}-going neutrinos. At Gran Sasso the low energy part of the flux peaks for arrival angles towards the west. For completeness we display the full simulation result, but for xenon and argon \cevns recoils, the relevant distribution is the lowest energy panel.

We do not have angular information for energies below 100 MeV, so we will extrapolate the angular distribution from the lowest energy bin of the HKKM2014 result. To understand how much this extrapolation might impact our results, in Fig.~\ref{fig:AtmFlux} we also showed $E^{\rm 90\%}_r$: the energy above which 90\% of events scatter, for a given initial neutrino energy. For LAr experimental thresholds around 30 keV, most of the possible neutrino energies are within the HKKM2014 spectrum and are thus safe to use their angular distributions. For LXe experimental thresholds in the range 2--10 keV, many more recoils will come from neutrinos in the lower energy component, so the extrapolation is a greater potential source of error. Fortunately, in the context of the directional sensitivity, this error should cause our results to be generally more conservative rather than less. The angular distribution is likely to become \emph{more} asymmetric towards lower energies because of the increasing importance of geomagnetic effects. So the angular distribution of recoils at low energies should be more anisotropic in reality than what we will assume: meaning greater potential discrimination between the WIMP and atmospheric neutrino signals could be possible. In any case, since the neutrino-nucleus scattering angles are large, we will integrate over azimuthal angles, and we will convolve our distributions with an effective angular resolution, this source of error will ultimately be quite minor. Since we have well-estimated energy spectra at these low energies and are accounting for a finite systematic uncertainty, the results for the non-directional sensitivity of xenon and argon experiments should be unaffected. However, as we will see, the sensitivity below the neutrino floor could be significantly enhanced if this uncertainty is reduced, either through further cosmic ray studies, or by direct measurements with neutrino experiments.

We do not have an analytic model describing $\drm^2\Phi/\drm E_\nu \drm \Omega_\nu$ for atmospheric neutrinos. So rather than evaluating Eq.(\ref{eq:nu_directionalrate}), we compute the directional event rate via a Monte Carlo simulation. We set up the simulation by first generating neutrino energies distributed as $E^2_\nu \, \drm\Phi_\nu/\drm E_\nu$, with zenith and azimuthal angles drawn from the distributions shown in Fig.~\ref{fig:AtmFlux_angle}. 
For a given recoil energy the \cevns cross section scales as $\drm R_\nu/\drm E_r \propto (1-E_r/E^{\rm max}_r)$. So for each $E_\nu$ we can generate a recoil energy from the correct linearly falling spectrum of recoil energies using $E_r = E^{\rm max}_r(1-\sqrt{u})$ where $u \in [0,1]$ is a number drawn from a uniform distribution. Then for each recoil energy we use the expression for $\cos{\beta}$ [Eq.(\ref{eq:kinematics})] to determine the nuclear scattering angle. The full recoil vector, $\qrhat$, is determined by deflecting the incoming neutrino direction, $\qnuhat$ by $\beta$ and rotating the new vector around the original direction by a uniformly sampled angle $\psi \in [0,2\pi)$ using Rodrigues' formula,
\begin{equation}
\qrhat  \rightarrow \qrhat \cos{\psi} + (\qnuhat \times \qrhat) \sin{\psi} + \qnuhat (\qnuhat \cdot \qrhat) (1-\cos{\psi}) \, .
\end{equation}
Finally, we can correct our distribution of recoil energies to account for the nuclear form factor suppression by calculating $F^2(E_r)$ for each recoil energy and discarding it with a probability $1-F^2(E_r)$. 

The resulting distribution of argon recoils between 2 and 100 keV is shown in Fig.~\ref{fig:AtmFlux_recoilangle}. The angular dependence of the incoming flux is largely washed out by the large nuclear scattering angles: $\langle \beta \rangle~=~$75$^\circ$ for the recoils in the energy window shown here. So rather than peaking at the horizon like the incoming flux, the event rate of nuclear recoils peaks weakly towards the zenith and nadir.

This distribution is the first stage of our background model. In principle, we should also include a seasonal variation in the integrated neutrino flux based on the atmospheric temperature, e.g.~Ref.~\cite{Tilav:2010hj}. The flux also modulates with the solar cycle. We have checked both of these modulations as reported in Ref.~\cite{Honda:2015fha}---on the total flux and on the angular and energy spectra---and find them to be negligible for our purposes. However, because the modulation is not known below 100 MeV we are relying on extrapolation. Nevertheless, any time dependence would certainly be smaller than the $\mathcal{O}(1)$ modulation of the WIMP signal's directional dependence due to the rotation of the Earth (see Sec.~\ref{sec:daily}), so it is safe to ignore.

\subsection{Other neutrino backgrounds}
Of the remaining neutrino backgrounds, the most relevant contributions to the nuclear recoil rate are from the highest energy solar neutrinos (\Boron and $hep$). The theoretical systematic uncertainties on the solar neutrino fluxes from standard solar models (SSMs) range from 0.6\% ($pp$ flux) to 30\% ($hep$ flux). In the case of the \Boron flux, the measurement uncertainty from a global analysis of neutrino data is the smaller uncertainty at 2\%~\cite{Bergstrom:2016cbh}, so we adopt this value instead.

The event rate for solar neutrinos as a function of recoil direction, energy and time is~\cite{O'Hare:2015mda},
\begin{align}\label{eq:solarnu}
  \frac{\drm^2 R_\nu(t)}{\drm E_r \drm \Omega_r} &=  \frac{1}{2\pi m_N}\left[ 1 + 2 e \cos\left(\frac{2\pi(t- t_\nu)}{T_\nu}\right) \right] \nonumber \\
   &\times\frac{\mathcal{E}(t)^2}{ E_\nu^\textrm{min}} \left(\dbd{\sigma}{E_r}  \dbd{\Phi}{E_\nu} \right)\bigg|_{E_\nu = \mathcal{E}(t)} \, ,
\end{align}
where,
\begin{equation}\label{eq:Eps}
  \frac{1}{\mathcal{E}(t)} = \frac{\hat{{\bf q}}_r \cdot \hat{{\bf q}}_\odot(t)}{E_\nu^\textrm{min}} - \frac{1}{m_N} \, .
\end{equation}
Due to the eccentricity of the Earth's orbit, $e = 0.016722$, the Earth-Sun distance has an annual variation ($T_{\nu} = 1$ year) leading to a modulation in the solar neutrino flux with a phase $t_{\nu} = 3$ days (relative to January 1). We have assumed that the flux is a delta function in the inverse of the direction towards the Sun, $\hat{{\bf q}}_\odot(t)$. The event rate is only non-zero for directions that satisfy $\cos^{-1}(\hat{{\bf q}}_r \cdot \hat{{\bf q}}_\odot(t)) < \pi/2$.

There is a small window of neutrino energies between 20--40 MeV where the dominant flux is from relic supernova neutrinos. The DSNB flux is the integral of the rate density for core-collapse supernova as a function of redshift, with the Fermi-Dirac neutrino spectrum expected from supernovae. Following Ref.~\cite{Beacom:2010kk} we take the flux to be the sum of distributions at the following temperatures for each neutrino flavour: $T_{\nu_e}~=~3$~MeV, $T_{\bar{\nu}_e}~=~5$~MeV and $T_{\nu_x}~=~8$~MeV, where $\nu_x$ represents the four remaining neutrino flavours. The DSNB is isotropic and constant over time so the directional CE$\nu$NS rate is simply,
\begin{equation}\label{eq:isonu}
\frac{\drm^2 R_\nu}{\drm E_r \drm \Omega_r} = \frac{1}{4\pi m_N} \int_{E_{\nu}^\textrm{min}} \dbd{\sigma}{E_r}  \dbd{\Phi}{E_\nu}  \drm E_\nu \,. 
\end{equation}

In Table~\ref{tab:neutrinos} we also listed the fluxes for reactor and geological antineutrinos for an experiment located at LNGS. Both produce recoil energies below a typical liquid xenon or argon detector threshold. They are unimportant in our main results but for completeness they are accounted for in our calculation of the neutrino floor in Sec.~\ref{sec:neutrinofloor}. For the reactor neutrino intensity we assume the fission fractions and average energy releases from Ref.~\cite{Ma:2012bm} combined with the spectra from Ref.~\cite{Mueller:2011nm}. We then find the flux by summing over all nearby nuclear reactors to LNGS~\cite{Baldoncini:2014vda}. For geoneutrinos we sum the radiogenic antineutrino spectra from U, Th and K isotopes which can be found in, for example Ref.~\cite{Ludhova:2013hna}, and then normalise to the flux at LNGS using the global reference model of Ref.~\cite{Huang:2013}. Similar calculations can be found in e.g. Refs.~\cite{Leyton:2017tza,Gelmini:2018gqa,Agostini:2019dbs}. Both reactor and geoneutrinos have interesting directional dependence but since they are not included in our final results we will not discuss them here.

\subsection{Dark matter scattering}\label{sec:DM}
As a final piece of theory input, we summarise our parameterisation of the WIMP signal and some assumptions we will make.

We can express the analogous event rate~$R_\chi$ for WIMP-nucleus elastic scattering events as~\cite{Gondolo:2002np,Drukier:1986tm},
\begin{equation}\label{eq:finaleventrate-directional}
   \frac{\drm^2 R_\chi(t)}{\drm E_r \drm\Omega_r} =\frac{1}{2\pi}\frac{\rho_0}{m_\chi} \frac{\mathcal{C} \sigma_{p} }{2 \mu_{\chi p}^2} F^2(E_r) \, \hat{f}(\vmin,\hat{\textbf{q}},t) \, ,
  \end{equation}
where $\mu_{\chi p}$ is the WIMP-proton reduced mass and $\sigma_p$ is the WIMP-proton scattering cross section. In this formula,
we have absorbed all the dependence on the nuclear content into the same form factor~$F(E_r)$ as introduced previously, and an ``enhancement factor''~$\mathcal{C}$ which is used so that we can extract the target-independent $\sigma_p$. For spin independent WIMP-nucleus scattering, the enhancement factor and cross section from Eq.(\ref{eq:finaleventrate-directional}) can be written,
\begin{equation}\label{eq:CSI}
\mathcal{C} \sigma_p \equiv \mathcal{C}^{\rm SI} \sigmapSI = | Z + (f_n/f_p)(A-Z)|^2 \sigmapSI \, ,
\end{equation}
when we have a nucleus with mass number $A$ and atomic number $Z$, and assign the couplings to neutrons and protons $f_n$ and $f_p$.

Throughout we will take the common assumption of equal couplings to protons and neutrons, $f_n/f_p = 1$, as generically found in models with a Higgs-like mediator~\cite{Hoferichter:2017olk}, though different values are possible~\cite{Frandsen:2011cg}. We only consider $\sigmapSI$ here because (1) it is the most commonly shown and most well-constrained cross section for liquid noble experiments; (2) it comes from the simplest WIMP-nucleus operator that produces a rate strongly mimicked by the neutrino background, and therefore has a neutrino floor; and (3) we would obtain largely similar qualitative conclusions as if we considered the full library of all possible low-energy operators (see e.g.~Ref.~\cite{Dent:2016iht} for a more detailed discussion of this).

Nevertheless, due to the wide variety of spin contents for different target nuclei, the neutrino floors for spin-dependent (SD) operators cover very different parts of well-motivated parameter space~\cite{Ruppin:2014bra}. Bubble chamber experiments such as PICO~\cite{Amole:2016pye, Amole:2017dex, Amole:2019coq} have set the tightest limits on SD-proton couplings, since they use targets involving $^{19}$F, which contains an unpaired proton. For DM models with SD-proton couplings around the neutrino floor~\cite{Beskidt:2017xsd}, bubble chambers, as well as directional TPCs using gases like SF$_6$~\cite{Vahsen:2020pzb} or CF$_4$~\cite{CYGNO:2019aqp}, may be more appropriate options than liquid noble experiments.

The final piece to consider is the velocity distribution of DM. We adopt the Gaussian standard halo model (SHM) for which the galactic frame distribution is,
\begin{equation}
\begin{aligned}
f_{\rm gal}(\mathbf{v})=\frac{1}{\left(2 \pi \sigma_{v}^{2}\right)^{3 / 2} N_{\mathrm{esc}}} & \exp \left(-\frac{|\mathbf{v}|^{2}}{2 \sigma_{v}^{2}}\right) \\
\times \Theta\left(v_{\mathrm{esc}}-|\mathbf{v}|\right) \, ,
\end{aligned}
\end{equation}
where,
\begin{equation}
N_{\mathrm{esc}}=\operatorname{erf}\left(\frac{v_{\mathrm{esc}}}{\sqrt{2} \sigma_{v}}\right)-\sqrt{\frac{2}{\pi}} \frac{v_{\mathrm{esc}}}{\sigma_{v}} \exp \left(-\frac{v_{\mathrm{esc}}^{2}}{2 \sigma_{v}^{2}}\right)
\end{equation}
and, $\sigma_v = 156~\kms$ and $\vesc = 533~\kms$. To obtain the laboratory frame distribution we need to apply a boost in velocity $\vlab \approx 240~\kms$ (see Ref.~\cite{Mayet:2016zxu} for how to calculate this in our lab-fixed coordinate system), which also is where the rate picks up a time dependence,
\begin{equation}
    f_{\rm lab}(\vbf,t) = f_{\rm gal}\big(\vbf+\vlab(t)\big) \, .
\end{equation}
In Eq.\eqref{eq:doublecrosssection} the lab-frame velocity distribution enters as its ``Radon transform''~\cite{Gondolo:2002np,Radon}:
\begin{equation}
   \hat{f}(\vmin,\hat{\textbf{q}},t) = \int \delta\left(\textbf{v} \cdot \hat{\textbf{q}} - \vmin\right) f_{\rm{lab}}(\textbf{v},t)\, \drm^3 v\, ,
\end{equation}
which is taken at $v_{\rm min}$, the minimum WIMP speed that can produce a recoil with energy $E_r$. We adopt this now slightly out of date SHM because we will compare our results to previous analyses. A discussion of more recent refinements made possible by data from the $\textit{Gaia}$ mission can be found in Ref.~\cite{Evans:2018bqy}.

\section{The neutrino floor}\label{sec:neutrinofloor}
\begin{figure*}
\begin{center}
\includegraphics[trim = 0mm 30mm 0mm 20mm, clip, width=0.9\textwidth]{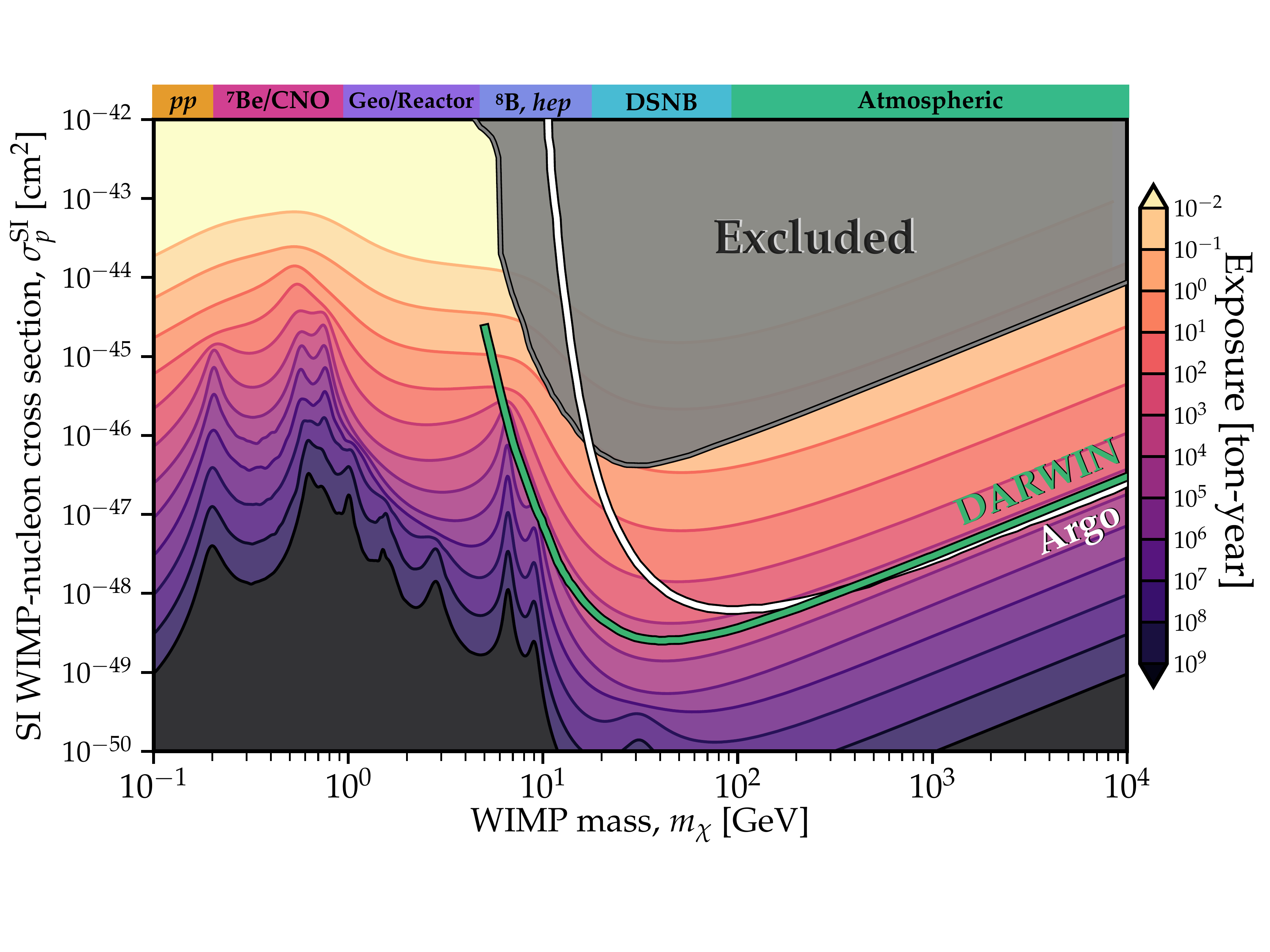}
\caption{Spin independent WIMP-proton cross section discovery limits which show the impact of every contribution to the neutrino background on an argon experiment. For completeness, we include components such as geoneutrinos and reactor neutrinos which would require impossibly large exposures and low recoil energy sensitivity to observe. The relevant source of neutrino for each range of masses is indicated on the upper horizontal axis. For comparison we show the projected reach of DARWIN~\cite{Aalbers:2016jon} and Argo~\cite{Aalseth:2017fik}. We caution that the range of exposures chosen here is only to illustrate the range of masses for which each neutrino background component is important. This range does not correspond to any realistic future experiment.} 
\label{fig:NuFloor}
\end{center}
\end{figure*} 

\begin{figure*}
\begin{center}
\includegraphics[trim = 0mm 0 0mm 0mm, clip, width=0.48\textwidth,angle=0]{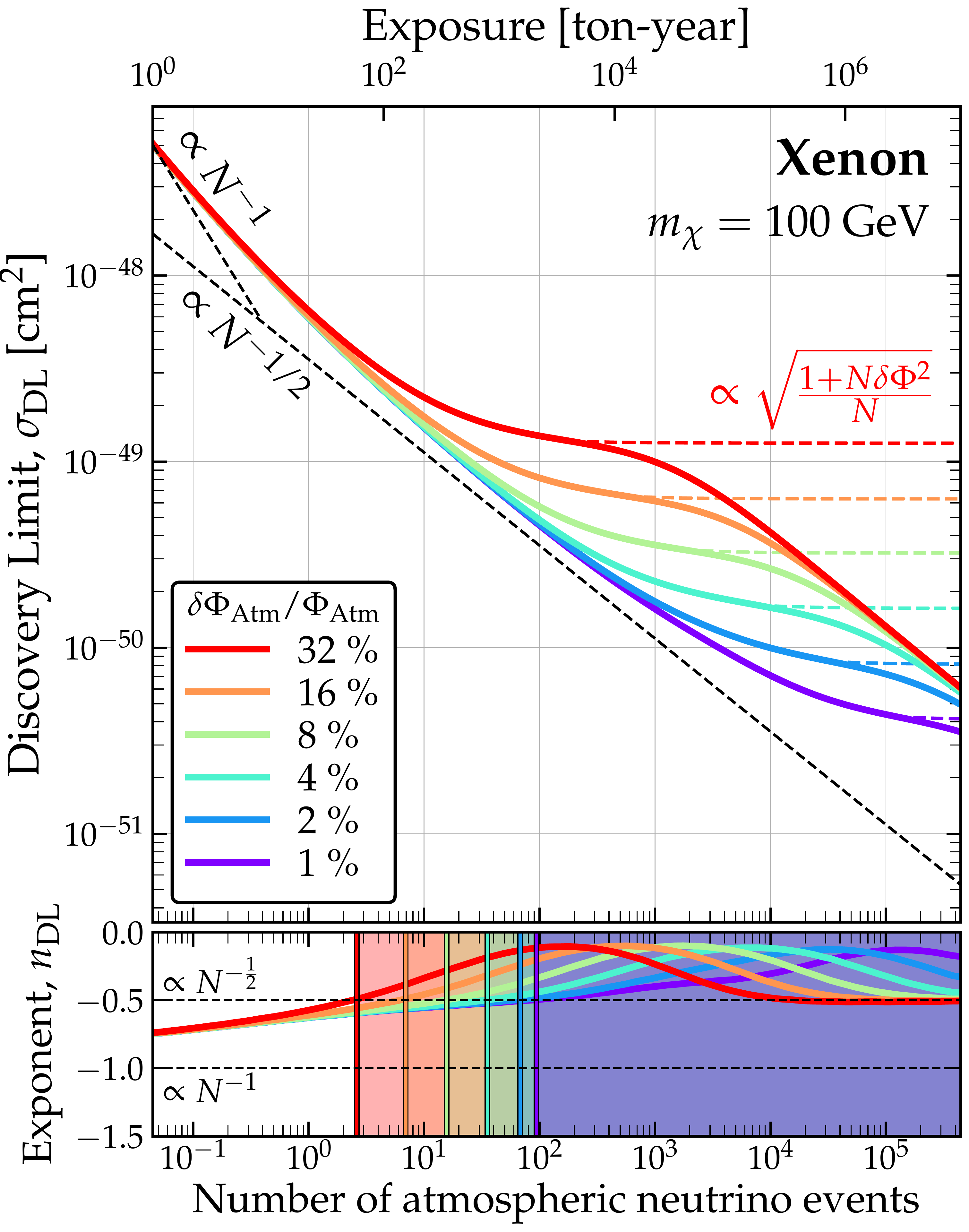}
\includegraphics[trim = 0mm 0 0mm 0mm, clip, width=0.48\textwidth,angle=0]{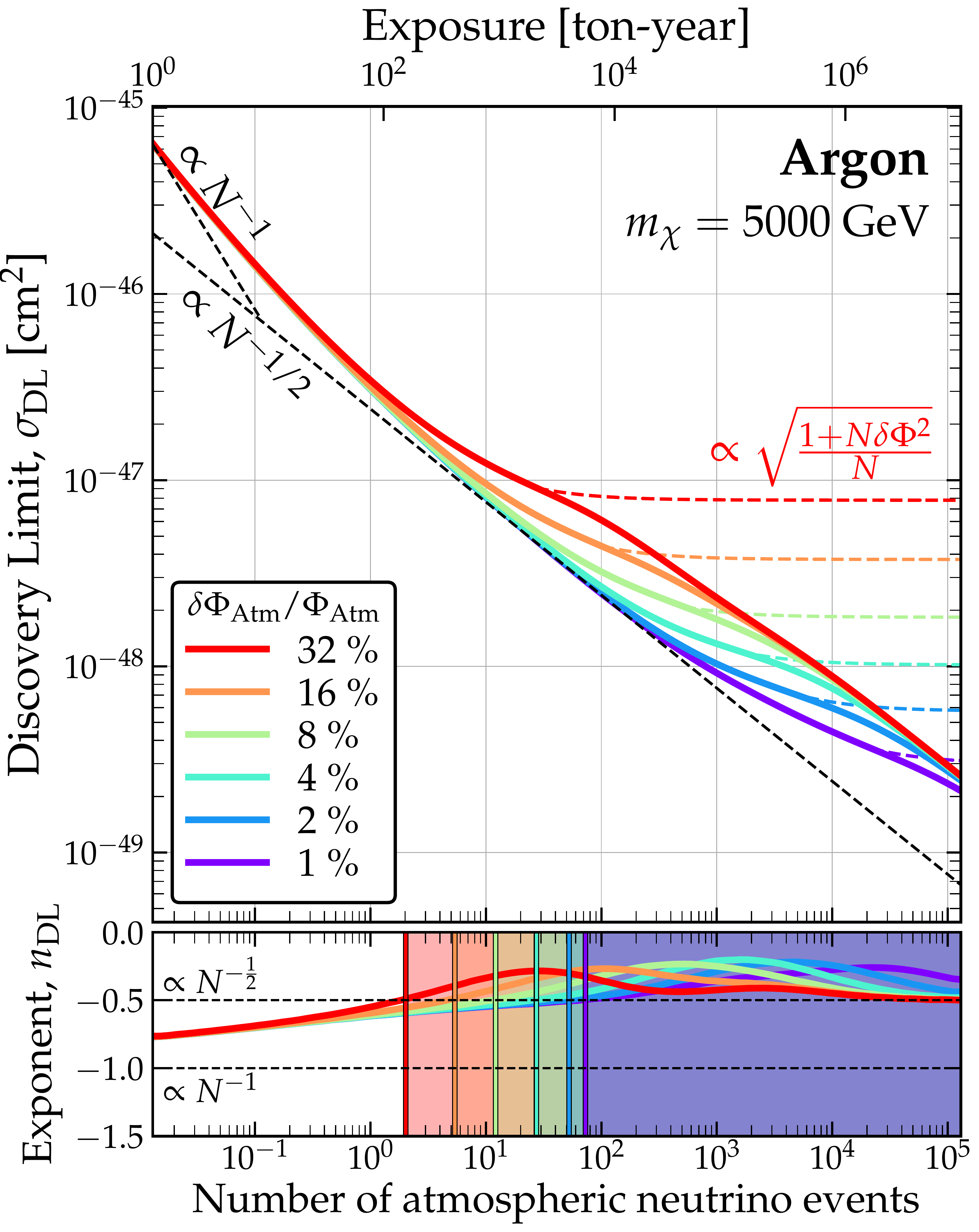}
\caption{Spin independent discovery limits at $m_\chi = 100$~GeV for a xenon target (left) and $m_\chi = 5000$~GeV for an argon target (right) as a function of the expected number of atmospheric \cevns events $N$, and the fractional uncertainty on the atmospheric neutrino flux, $\delta \Phi_{\rm Atm}/\Phi_{\rm Atm}$. We also indicate the three scaling regimes as a function of $N$ with dashed lines: (1) ``background-free'' $\sigma \sim N^{-1}$, (2) Poissonian $\sigma \sim N^{-1/2}$ and (3) saturation $\sigma \sim \sqrt{(1+\delta\Phi^2 N)/N}$. The bottom panels in each case show the logarithmic scaling exponent defined as: $n_{\rm DL} \equiv \drm \ln\sigma_{\rm DL}/\drm \ln N$. The two different masses that have been chosen here correspond to the cases where the recoil distributions for xenon and argon most strongly overlap with the \Boron recoil rates. These choices therefore correspond to the masses for which the discovery limits have the largest departure from the Poissonian scaling regime.}
\label{fig:nufluxunc}
\end{center}
\end{figure*} 
\subsection{Statistical methodology}
We quantify the impact of the neutrino background on the detection of a WIMP signal in terms of discovery limits, so we will first briefly summarise how to compute them using the profile likelihood ratio test~\cite{Cowan:2010js}.

Our parameters of interest are the WIMP's mass and some cross section: $m_\chi$ and $\sigma$. The parameters controlling the background are the neutrino flux normalisations $\boldsymbol{\Phi} = \{ \Phi^1, ..., \Phi^{n_\nu} \}$, for $n_\nu$ neutrino species. We use a binned likelihood written as the product of the Poisson probability distribution function ($\mathscr{P}$) in each bin, multiplied by Gaussian likelihood functions for the uncertainties on each neutrino flux normalisation ($\mathscr{G}(\Phi^j)$):
\begin{align}\label{eq:likelihood}
 \mathscr{L}(m_\chi,\sigma,\boldsymbol{\Phi}|\mathcal{M}) &= \prod_{i=1}^{N_\textrm{bins}} \mathscr{P} \left(N_\textrm{obs}^i \bigg| N^i_\chi + \sum_{j=1}^{n_\nu} N^{i}_\nu(\phi^j)\right)  \nonumber \\
&\times \prod_{j=1}^{n_\nu} \mathscr{G}(\Phi^j) \, .
\end{align}
The Gaussian distributions have mean values $\Phi^j$ and standard deviations $\delta \Phi^j$, as listed in Table~\ref{tab:neutrinos}. The Poisson probabilities at the $i$th bin are taken for an observed number of events $N_\textrm{obs}^i$, given an expected number of WIMP events $N_\chi^i$ and the sum of the expected number of neutrino events for each neutrino species $N_\nu^i(\Phi^j)$. The way in which the space of observables is partitioned into bins will depend on the type of experiment in question. For the neutrino floor we assume that the only observable is recoil energy, $E_r$. In Sec.~\ref{sec:CRmodel} when we introduce directional sensitivity we will assume a multidimensional binning over observables which involve energy, angle and time.

The profile likelihood ratio test compares the WIMP-less, background-only model $\mathcal{M}_{\sigma=0}$ with parameters $(\sigma=0,\boldsymbol{\Phi})$ against the WIMP+background model $\mathcal{M}$ with parameters $(\sigma, \boldsymbol{\Phi})$. Since the background-only model is insensitive to $m_\chi$, the typical procedure involves fixing the mass but repeating the test over a range of values to map the discovery limit. The two models then only differ by one parameter, $\sigma$. To test for $\sigma > 0$ we write down the ratio of the two maximised likelihood functions (now ignoring $m_\chi$),
\begin{equation}\label{eq:likelihood-ratio}
\Lambda = \frac{ \mathscr{L} (0,\hat{\hat{\boldsymbol{\Phi}}} | \mathcal{M}_{\sigma=0}) }{\mathscr{L} (\hat{\sigma},\hat{\boldsymbol{\Phi}} | \mathcal{M})  }\, ,
\end{equation}
where~$\mathscr{L}$ is maximised at $\hat{\hat{\boldsymbol{\Phi}}}$ when $\sigma$ is set to 0, and $(\hat{\sigma},\hat{\boldsymbol{\Phi}})$ when $\sigma$ is a free parameter. Our null hypothesis is the WIMP-less model $\mathcal{M}_{\sigma=0}$, which is a subset of the more general model $\mathcal{M}$.

We define the test statistic for this likelihood ratio as,
\begin{equation}
	q_0 = \left\{ \begin{array}{rl}
	-2\ln \Lambda  & \, \,  \hat{\sigma}>0  \,,\\
	0  & \, \, \hat{\sigma}\le 0, \, \,  \,.
	\end{array} \right. 
\end{equation}
Since the two models differ by the fixing of one parameter, and our null hypothesis has a parameter set to the boundary of the allowed space, Chernoff's theorem~\cite{Chernoff:1954eli} holds. This is a generalisation of Wilk's theorem and states that the statistic $q_0$ is asymptotically distributed according to $\frac{1}{2}\chi^2_{1}+\frac{1}{2}\delta(0)$ when the $\mathcal{M}_{\sigma=0}$ hypothesis is true. The practical consequence of this for us is that the significance of the WIMP signal tested against the background-only hypothesis is simply $\sqrt{q_0}$. See Ref.~\cite{Cowan:2010js} for a detailed discussion of the use of these asymptotic formulae. We therefore define a 3$\sigma$ discovery limit at some $P\%$ confidence level (C.L.) to be the minimum value of $\sigma$ for which $P\%$ of the asymptotic distribution of $\sqrt{q_0}$ is greater than 3.

The distribution of $q_0$ under the model $\mathcal{M}$ would normally be calculated using many Monte Carlo realisations of pseudodata. There is however a trick we can use to greatly reduce this computational expense. We can instead instantly calculate the median discovery limit using the Asimov dataset~\cite{Cowan:2010js}. This is a hypothetical scenario in which the observation exactly matches the expectation for a given model, i.e. $N^i_{\rm obs} = N^i_{\rm exp}$ for all $i$. It can be shown that the test statistic computed assuming this dataset asymptotes towards the median of the model's $q_0$ distribution as the number of observations increases~\cite{Cowan:2010js}. In analyses such as ours this turns out to be an extremely good approximation. Henceforth, all of our limits are defined as 3$\sigma$ discovery limits at 50\% C.L. This is a mild departure from some previous work on this subject---e.g. Refs.~\cite{Billard:2013qya,OHare:2016pjy} which used discovery limits at 90\% C.L.---but overall is a minor quantitative difference, one which is worth the computational saving.

\subsection{Impact of the neutrino background}
The impact of neutrinos on the discovery of DM depends on the size of the neutrino background and---though not often stated explicitly---its systematic uncertainty. A feeble WIMP signal is saturated not just when the number of signal events is simply less than the background, but when that excess of events is smaller than the potential statistical \emph{fluctuation} in the background. 

More precisely, as the exposure $\mathcal{E}$ of an experiment increases, the background grows linearly $\sim\mathcal{E}$ but the number of events required to detect the WIMP at a fixed significance should only grow with $\sim\sqrt{\mathcal{E}}$, for Poissonian statistics. But eventually the exposure will be large enough that $\sqrt{\mathcal{E}}/\mathcal{E} < \delta \Phi$, where $\delta \Phi$ is the uncertainty on the background. At this point the WIMP signal, which would have been detectable otherwise, only provides excess events at a lower level than the expected statistical fluctuation. If there is no other way to distinguish the WIMP events from background, the minimum discoverable cross section will plateau for increasing $\mathcal{E}$. In practice though, recoil energy information provides a weak discriminant, so this saturation only occurs strongly when the range of recoil energies for certain WIMP masses closely overlap with the spectrum of a particular component of the neutrino background.


Figure~\ref{fig:NuFloor} shows those WIMP masses which are most impacted by each component of the neutrino background listed in Table~\ref{tab:neutrinos}. The discovery limits in this case correspond to arbitrarily large and sensitive experiments: the full range of WIMP masses and cross sections shown here is demonstrably not accessible to any single experiment. Rather, this plot serves to illustrate the ranges of WIMP models where each neutrino background is important. For consistency we choose an argon target nucleus here, but equivalent plots for other nuclei look similar to this.

The focus area for this study are masses above $m_\chi \sim 10$ GeV. For argon-based experiments, the sensitivity starts to be impacted by the atmospheric neutrino background for $m_\chi \gtrsim 100$~\GeV and $\sigmapSI \sim 10^{-(48\textrm{--}49)}$~cm$^2$. Reaching these values requires exposures $\gtrsim$100 ton-year, for example DarkSide-20k, DARWIN, or Argo. For xenon-based experiments, the range of WIMP masses and cross sections is roughly similar but the exposures needed are slightly smaller due to the $A^2$ scaling of the WIMP-nucleus cross section. To see this in more detail, the left and right panels of Fig.~\ref{fig:nufluxunc} show the discovery limit $\sigma_{\rm DL}$ as a function of number of atmospheric neutrino events, at two fixed masses.

The scaling of $\sigma_{\rm DL}$  evolves through three regimes for increasing numbers of background events, $N$. Initially, for $N<1$ the limit approaches a $1 \propto 1/N$ scaling; as is the case for experiments that are effectively background-free, i.e. have less than one expected background event in their exposure. Then as $N$ increases, the limit transitions into a standard Poissonian background subtraction regime $\propto 1/\sqrt{N}$. Eventually the WIMP signal is hidden beneath the potential neutrino background fluctuation, controlled by $\delta \Phi$ and the discovery limit briefly follows~\cite{Billard:2013qya},
\begin{equation}
\sigma_{\rm DL} \propto \sqrt{\frac{1+N\delta \Phi^2}{N}} \, .
\end{equation}
In Fig.~\ref{fig:nufluxunc}, this regime takes over around $N~\gtrsim 100$ for the value of $\delta \Phi = 25\%$ (which is our baseline assumption for later results). If the WIMP signal and \cevns background were identical, this regime would persist for arbitrarily large $N$. The discovery limit would plateau and never improve. However for both xenon and argon there is never a value of $m_\chi$ for which the \cevns background perfectly matches the WIMP signal. Eventually there will be enough statistics to distinguish between the spectra. The scaling of $\sigma_{\rm DL}$ once it breaks past this third scaling regime returns to Poissonian background subtraction $\propto N^{-1/2}$, but only after $\sim 10^4$ background events are collected. The limit is thus not truly a hard floor, but a soft barrier.

The ``softness'' of the neutrino floor can be understood by considering the lower panels of Fig.~\ref{fig:nufluxunc}. In these we show the exponent of the gradient of the discovery limit versus exposure, i.e. $n_{\rm DL} \equiv \drm \ln\sigma_{\rm DL}/\drm \ln N$ For xenon this number almost reaches zero because the recoil spectra of a 100 GeV WIMP and atmospheric neutrinos are very similar. For argon they are less similar, and the exponent only reaches $-0.25$. In other words, the argon floor is softer and less problematic than the xenon floor usually shown alongside WIMP direct detection results. This fact seems to have not been stated straightforwardly in the literature before, despite the fact that it can be gleaned from previous work~\cite{Ruppin:2014bra}.

\section{Directional detection}\label{sec:directionality}
In Fig.~\ref{fig:nufluxunc} we saw that the neutrino background is eventually overcome because $\drm R_\chi/\drm E_r$ and $\drm R_\nu/\drm E_r$ are slightly different. In a similar fashion, any further discriminating information will help subtract the background further. This information could be from the annual modulation signals~\cite{Davis:2014ama} or from target nucleus dependence~\cite{Ruppin:2014bra} for example. In the context of atmospheric neutrinos, the low energy flux would also be subject to an 11 year modulation due to the solar cycle which could also be used as a discriminant for long exposure experiments.

However the most powerful discriminant of neutrinos against DM-induced recoils is their directionality. In fact, when the dominant background is from solar neutrinos, experiments with good 3-d track reconstruction could achieve discovery limits that scale even steeper than Poissonian background subtraction~\cite{O'Hare:2015mda}. This is because the DM wind and the Sun never coincide on the sky: there are regions of signal space in energy, angle and time that are guaranteed to have low numbers of expected \Boron recoil events. 


The most developed method of directional detection involves gas-based TPCs in which recoil directions can be observed directly. However the low inherent target masses of gas targets make this technique only appropriate for low masses and low energy thresholds. Directional signals in much larger solid or liquid-state experiments are therefore particularly desirable, but the very short track lengths for \keV recoils in high-density media make the direct measurement of tracks extremely difficult. So instead we seek methods of obtaining directional sensitivity indirectly via other observables. In certain anisotropic scintillators like ZnWO$_4$ and stilbene for example the scintillation yield and pulse shape can depend on the orientation of the recoil event with respect to the crystal axes~\cite{Belli:1992zb,Spooner:1996gr,Shimizu:2002ik,Cappella:2013rua,Sekiya:2003wf}. Such an experiment would exploit the rotation of the Earth and attempt to observe modulations in the detector response for events collected at different times during the day. So rather than directly reconstructing a 3-d angular distribution, directionality would be inferred from the characteristic phase and amplitude of certain daily modulations.

\subsection{Columnar recombination}
An indirect measure of directionality called columnar recombination may be present in xenon or argon experiments. The effect appears when the recombination of a cloud of electrons and ions depends upon the direction of an applied electric field. To gain a rough picture, we can use the Onsager geminate theory~\cite{Onsager:1938zz} which assumes that electrons reattach to ions within a radius $r_{\rm O} = e^2/4\pi\varepsilon E_e$: when the ion's Coulomb potential overcomes the electron's energy $E_e$ ($\varepsilon$ is the dielectric constant of the medium). When a primary ionisation cloud generated by a recoil event is drifted with an electric field, some of the ions and electrons will recombine. The amount of ionisation that is ultimately detected from the event may therefore be dependent on the angle between the straggled recoil track and the electric field. Specifically, we expect to detect less ionisation when the recoil track is parallel to the field because the electrons must drift through the ionisation cloud, giving them a higher chance of recombining. Ideally, fluorescence from the recombining of the electrons and ions would also be observable as an additional scintillation signal. In this case, directionality would be encoded in the form of an asymmetry in the ratio of scintillation and ionisation yields for tracks parallel and perpendicular to the drift field. Parallel tracks would produce more scintillation, and perpendicular tracks, more ionisation.

The first experimental study of columnar recombination dates back to 1913 by Jaff\'e~\cite{Jaffe:1913} but was only suggested to be of potential use in DM experiments in 2013 by Nygren~\cite{Nygren:2013nda}. Subsequently, the effect has been investigated experimentally using $\alpha$ tracks in high-pressure xenon gas~\cite{MuAaoz:2014uxa}, 50--250 MeV proton tracks in LAr by ArgoNeuT~\cite{Acciarri:2009xj}, and nuclear recoils in LAr by SCENE~\cite{Cao:2014gns}.

The size of the recombination effect can be roughly parameterised using the aspect track aspect ratio $L/r_{\rm O}$. In LXe, a 30 keV track length is approximately 35 nm, so are on the same scale as, or smaller than, the Onsager radius of $r_{\rm O} \approx 54$~nm. This means the recombination aspect ratio is sadly always $<1$ and the only hope for observing the effect in xenon is in the gas phase. Nygren~\cite{Nygren:2013nda} suggested that 10 bars of high pressure xenon gas could allow the effect to be measured at an acceptably low threshold. This would need a $\sim$20~m$^3$  volume per ton of xenon.

In LAr the Onsager radius is slightly larger $r_{\rm O} \simeq 80$~nm, but so are the tracks, e.g. $L\simeq 90$~nm for a 36~keV recoil and 135~nm for a 57 keV recoil. The directional asymmetry between the scintillation yield of neutron-induced nuclear recoils in liquid argon was tested at these energies~\cite{Cao:2014gns}. The measured asymmetry of around $\lesssim$0.95 is only statistically significant for their 57 keV beam energy, as one might expect given this simple argument for the energy required to generate columnar recombination. Unfortunately, the corresponding asymmetry in the ionisation signal was not observed. The Recoil Directionality (ReD) experiment as a part of DarkSide is now investigating this further with a small scale liquid argon TPC.

One important way in which columnar recombination could be made more detectable is with the inclusion of trimethylamine (TMA) or triethylamine (TEA). These dopants help in a number of ways. Firstly the noble gas-TMA/TEA mixtures can exploit Penning transfer in which excited atoms of xenon or argon can de-excite via ionising a molecule of TMA or TEA. This means that the part of the recoil energy that would be lost due to excitations can be converted into additional ionisation and therefore increased potential recombination. Secondly, the dopants have large inelastic cross sections at low energy as well as a large number of vibrational and rotational modes meaning it can help hasten the thermalisation of the drifting electrons: minimising diffusion and enhancing directionality. For example the addition of TMA has been shown to reduce and maintain the diffusion of electrons in microphysics simulations of high-pressure xenon within 2$\upmu$m well after 0.1 ns~\cite{Nakajima:2015dva}, which is the typical size of a 30 keV recoil for that gas density.

\subsection{Daily modulation}\label{sec:daily}
\begin{figure}
\begin{center}
\includegraphics[trim = 0mm 0 0mm 0mm, clip, width=0.49\textwidth]{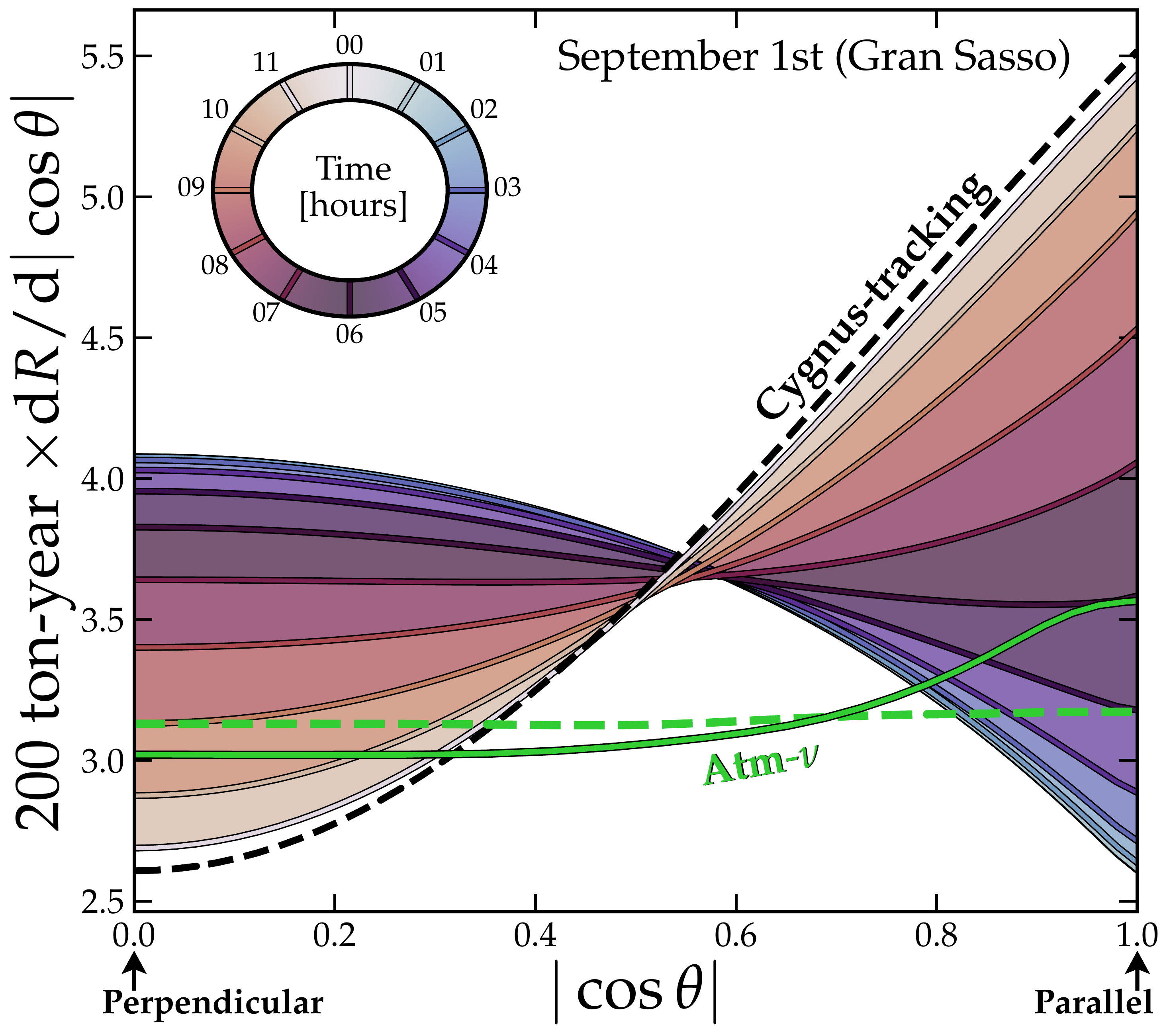}
\caption{One half-day evolution of the absolute value of the zenith angle $|\cos{\theta}|$, for a 5000 GeV WIMP scattering with argon, and integrated above $\sim$30 keV. For a drift field aligned vertically, $\cos{\theta} = 0$ corresponds to perpendicular tracks (minimal columnar recombination) and $\cos{\theta} = \pm 1$ corresponds to parallel tracks (maximal columnar recombination). The SI cross section is $\sigmapSI = 2.5\times 10^{-48}$~cm$^2$. We show only half a day because the distributions for 12:00--23:00 are roughly the same. The distributions of atmospheric neutrino recoils are shown in green. For both the atmospheric and WIMP distributions we show the Cygnus-tracking case with dashed lines and the stationary detector case with solid lines.} 
\label{fig:daily1d}
\end{center}
\end{figure} 
Columnar recombination is dependent on the angle with respect to the drift direction, $\cos{\theta}$, but since the effect has no head/tail signature, only the axial angle $|\cos{\theta}|$ can be inferred. Only having access to a one-dimensional projection of the full angular distribution is problematic for background rejection since there is limited room for the signal and background distributions to differ. So the daily modulation of $|\cos{\theta}|$ brought about by rotation of the Earth is essential to make sense of the signal.

In Fig.~\ref{fig:daily1d} we show the expected distribution of $|\cos{\theta}|$ for a 5000 GeV WIMP and the atmospheric neutrino background in an argon experiment. The daily modulation shown corresponds to an experiment located at LNGS with latitude and longitude ($42.5^\circ$,~$13.6^\circ$), on September 1st. We show only the distribution as a function of $|\cos{\theta}|$ by integrating over $E_r$. We weight the energy spectrum by a nuclear recoil acceptance function taken from the DarkSide projection (Fig.~92 of Ref.~\cite{Aalseth:2017fik}), which gives us a threshold of $\sim 30$~keV. For later results involving xenon we use the projection for LZ~\cite{Akerib:2018lyp}, which sets a $\sim$4 keV threshold.

The distributions of $|\cos{\theta}|$ show some distinction between the signal and the background, but both depend rather weakly on direction. A major reason for why we have lost so much directionality is because of the lack of a head/tail signature. The WIMP signal is a dipole, so being unable to recognise the forward/backward sense of individual events is important for maximising the anisotropy. It is well worth thinking how we could try to compensate for this loss of information. In the original paper proposing columnar recombination for a DM search~\cite{Nygren:2013nda} it was suggested that a head-tail signature could be obtained via the differences in the modulations of two stacked TPCs that would be tilted at the correct angle so that they align---one parallel and the other antiparallel---with Cygnus once per day. This would not work because the recombination signal is dependent on $|\cos{\theta}|$: the modulation signals in both stacked TPCs would be the same. If a double-TPC design was to be used, it would be better to align them orthogonally, where once per day one would be aligned and the other perpendicular. But under this configuration the TPCs would not be able to share a cathode, which was an attractive feature of the stacked design.

\begin{figure*}[tbp]
\begin{center}
\includegraphics[trim = 0mm 0 0mm 0mm, clip, width=0.98\textwidth]{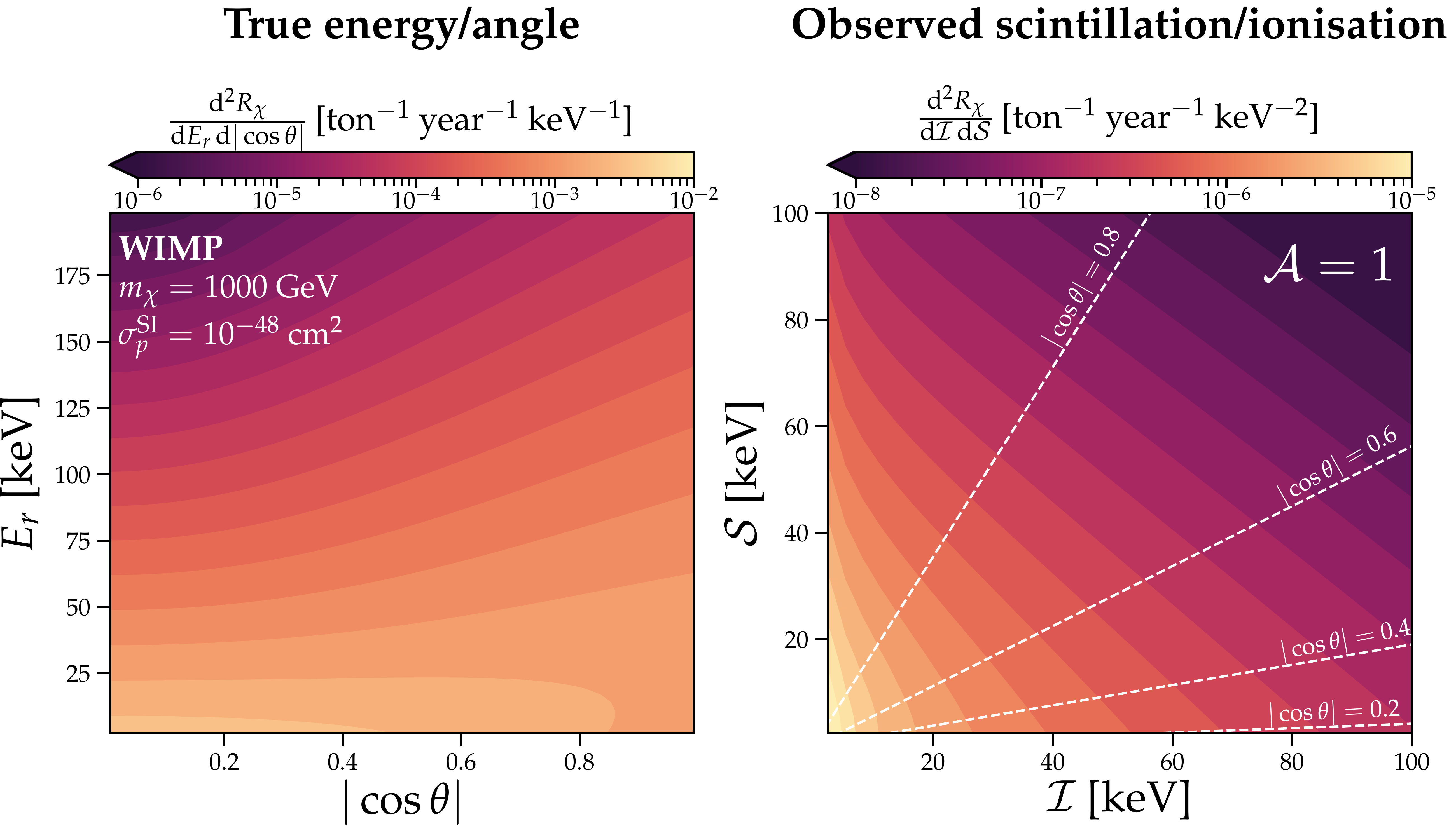}
\includegraphics[trim = 0mm 0 0mm 0mm, clip, width=0.98\textwidth]{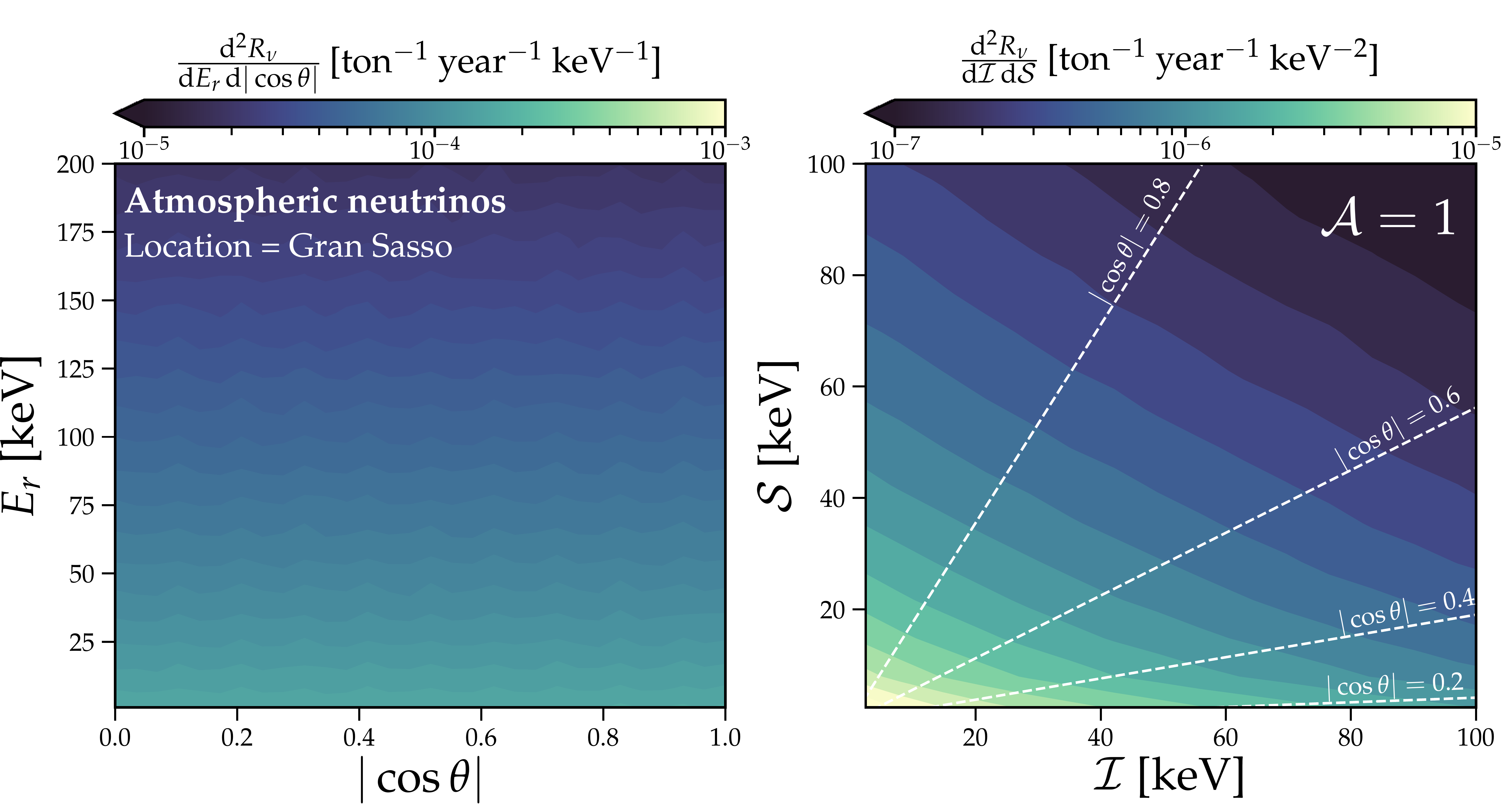}
\caption{WIMP signal (top panels) and atmospheric neutrino background distributions (bottom panels) for an argon target. In both cases the left-hand panels show the underlying distribution of argon recoil energies and angles, and the right-hand panels show the resulting distribution of the measurable quantities of scintillation and ionisation energy, calculated using Eq.(\ref{eq:SI}). We take the WIMP distribution at the time when the angle between the drift axis and Cygnus is minimised (which also corresponds to our ``Cygnus-tracking'' benchmark from Fig.~\ref{fig:daily1d}). We show contours of constant $|\cos{\theta}|$ as dashed lines.} 
\label{fig:recombinationsignal}
\end{center}
\end{figure*} 

\begin{figure}[tbp]
\begin{center}
\includegraphics[trim = 0mm 0 0mm 0mm, clip, width=0.49\textwidth]{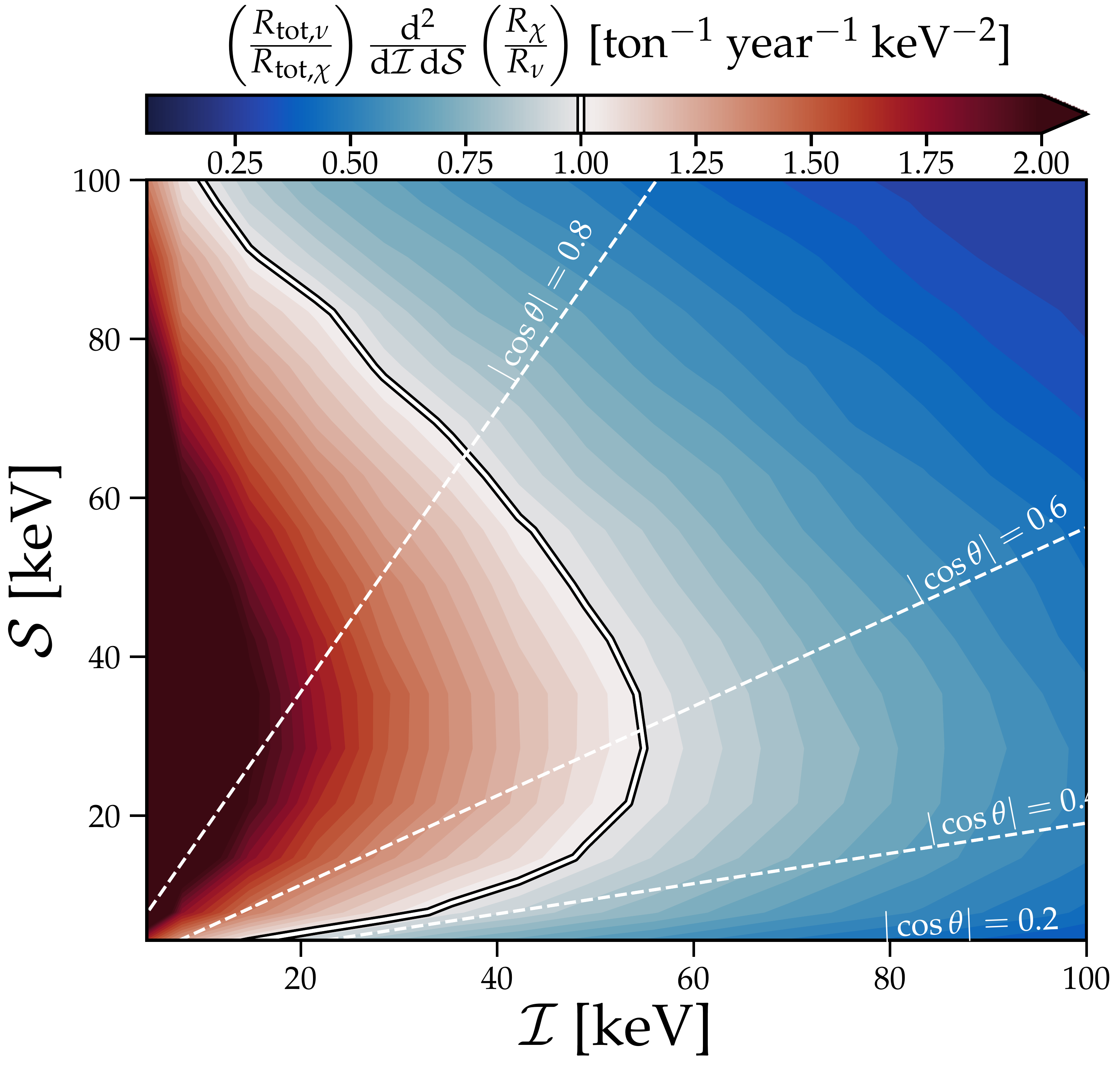}
\caption{The contrast between the WIMP and neutrino $(\mathcal{I},\mathcal{S})$ distributions from Fig.~\ref{fig:recombinationsignal}. We display the ratio between the distributions $R_\chi/R_\nu$, after both have been normalised by the total integrated rate $R_{\textrm{tot},\chi}$ and $R_{\textrm{tot},\nu}$.  The white line partitions this ratio of distributions between ranges of energies for which the WIMP rate is higher (red) and energies where the neutrino rate is higher (blue).
\label{fig:recombinationsignal_contrast}}
\end{center}
\end{figure} 

Thinking about potential alternatives, the most well-optimised orientation would be a ``Cygnus-tracking'' detector, i.e.~one which always points towards the direction of the DM wind: $\hat{\mathbf{v}}_\mathrm{lab}$. In this configuration the $|\cos{\theta}|$ distribution would remain constant over the day, and be maximally anisotropic. We show this distribution as a black dashed line in Fig.~\ref{fig:daily1d}. The WIMP signal for a Cygnus tracking experiment would always be peaked towards $|\cos{\theta}| = 1$. In the stationary mode (shown by solid lines in Fig.~\ref{fig:daily1d}, the distributions of $|\cos{\theta}|$ come very close to the Cygnus-tracking distribution around once per day (at around 0 hours on the date shown here). This occurs because we have located the experiment at Gran Sasso, where the zenith does happen to roughly align with the WIMP wind around once a day. In all locations with latitudes in the range $\pm(41^\circ$--$51^\circ)$ this will occur (which turns out to be the case for most underground labs).

A Cygnus-tracking experiment would require the target, detector, and any necessary shielding to all be mounted in some way on an equatorial telescope, which would slowly rotate over the day and night. This setup is not entirely without precedent, the nuclear emulsion-based directional detector NEWSdm is proposed to operate in this way~\cite{Aleksandrov:2016fyr,Agafonova:2017ajg}. For them Cygnus-tracking is essential, because the experiment is time-integrated: the emulsion plates need to be removed from the detector and analysed with a nano-imaging tracker in order for the nm-scale tracks to be identified~\cite{Natsume:2007zz,Naka:2013mwa}. The lack of Cygnus-tracking would reduce their sensitivity to WIMPs by a factor of around 1.5--3 under an isotropic background~\cite{OHare:2017rag}. The equatorial telescope design is already a considerable layer of added complexity for NEWSdm. For a much larger dual-phase noble detector, this complexity would probably be prohibitive. Here we simply take the scenario of ``Cygnus-tracking'' as an optimal strategy because it may turn out that it is required in some form to realistically make use of the columnar recombination signal.

\subsection{A simple model for columnar recombination}\label{sec:CRmodel}
While there are two notable historical models to describe columnar recombination---the original columnar model of Jaff\'e~\cite{Jaffe:1913} and the box model of Thomas and Imel~\cite{Thomas:1987zz}---both failed to capture the observed behaviour of proton tracks in liquid argon seen by ArgoNeuT~\cite{Acciarri:2009xj}. A more sophisticated model based on an elongated ellipsoidal shape for the initial ionisation distribution performed better~\cite{Cataudella:2017kcf}, but will need further experimental data to test its validity for nuclear recoils. 

Given the paucity of experimental data, there is little point in attempting a fully realistic model for columnar recombination of nuclear recoil tracks in high pressure gaseous xenon or in liquid argon. So our goal here is instead framed around asking how strong a directional signal in a similar format to columnar recombination would be required to search below the neutrino floor. In order to not drastically overestimate the capabilities of future detectors, we should account for some inevitable limitations of using the effect in practice, given that the directionality is inferred, rather than directly measured, through ionisation ($\mathcal{I}$) and scintillation ($\mathcal{S}$) signals.

We model columnar recombination by enforcing the measured scintillation and ionisation energies to be dependent on $|\cos{\theta}|$. Reference~\cite{MuAaoz:2014uxa} found that the recombination inferred via collected ionisation in a xenon+TMA mixture approximately scales with $\cos^2{\theta}$. Given this result, and that the more complex parameterisation of Ref.~\cite{Cataudella:2017kcf} for proton tracks in LAr, also resembles a $\cos^2{\theta}$ scaling, we will adopt this as a preliminary approximation. A prior theoretical study~\cite{Mohlabeng:2015efa} also implemented a toy model that scales in this way. Ours is similar but we will introduce an extra parameter $\mathcal{A}$ to allow us to tune the strength of the directional asymmetry. The scintillation and ionisation yields follow:
\begin{align}\label{eq:SI}
\mathcal{I}(E_r, \cos{\theta}) & = \epsilon_\mathcal{I}(E_r) E_r(1-\mathcal{A}\cos^2{\theta}) \nonumber \\
\mathcal{S}(E_r,  \cos{\theta}) & =  \epsilon_\mathcal{S}(E_r) E_r\mathcal{A}\cos^2{\theta} \, , 
\end{align}
where $\epsilon_{\mathcal{I},\mathcal{S}}$ are some efficiency functions for the measurement of ionisation/scintillation which we will take as a function of true recoil energy. 

The case $\mathcal{A}\rightarrow 1$ corresponds to the idealised limit of a maximal columnar recombination: when all the measurable energy is converted into recombination (scintillation) when the track is parallel to the drift field, and none of the electrons recombine when the track is perpendicular. The ionisation signal depends on the remaining energy after recombination, hence the $1-\mathcal{A}\cos^2{\theta}$ scaling. For both WIMPs and neutrinos we calculate the recoil distributions as a function of $(\mathcal{I},\mathcal{S})$ and account for a finite energy resolutions by smoothing the 2d distribution with a Gaussian kernel with widths $(\sigma_{\rm S1},\sigma_{\rm S2})$. We take these two energy resolutions along with the efficiency functions (which are all energy dependent) from simulation results which can be found in, for example, Ref.~\cite{Schumann:2015cpa} for xenon, and Ref.~\cite{Aalseth:2017fik} for argon.

The precise details of this model will inevitably need to be expanded upon in light of future experimental data. One obvious simplification we have made is to take $\mathcal{A}$ as constant in energy, whereas in reality it will be $\sim 0$ below some threshold and increase with energy. The threshold for columnar recombination will approximately correspond to when the track's aspect ratio $L/r_{\rm O}$ drops below 1. In LAr this occurs at $\sim 30$~keV and SCENE observed a $\sim95\%$ scintillation yield relative to zero electric field for $E_r = 57$~keV. So for a threshold of $\sim 30$ keV, a value of $\mathcal{A} \sim 0.05$ would approximately capture this experimental result. Though we stress caution here as this value depends on the choice of electric field strength, and the effect was not observed in the ionisation yield.

In Fig.~\ref{fig:recombinationsignal} we compare the \emph{underlying} WIMP and neutrino $(E_r,|\cos{\theta}|)$ distributions with the \emph{measurable} $(\mathcal{I},\mathcal{S})$. For clarity we show semi-idealised results here, before applying the scintillation or ionisation energy resolutions and efficiencies. This is so that the mapping of the underlying variables to the measured variables is made clearer with contours of constant $\cos{\theta}$. In the following results however all the above effects are accounted for. Figure~\ref{fig:recombinationsignal} effectively represents the signal and background models that the results in the next section are based upon. For added clarity we show the ratio of the normalised WIMP and neutrino event rates as a function of $(\mathcal{I},\mathcal{S})$ in Fig.~\ref{fig:recombinationsignal_contrast}. This figure allows us to see the energy ranges that are relatively dominated by neutrinos or WIMPs.

\section{Results}\label{sec:results}
\begin{figure*}[tbp]
\begin{center}
\includegraphics[trim = 0mm 0 0mm 0mm, clip, width=0.49\textwidth]{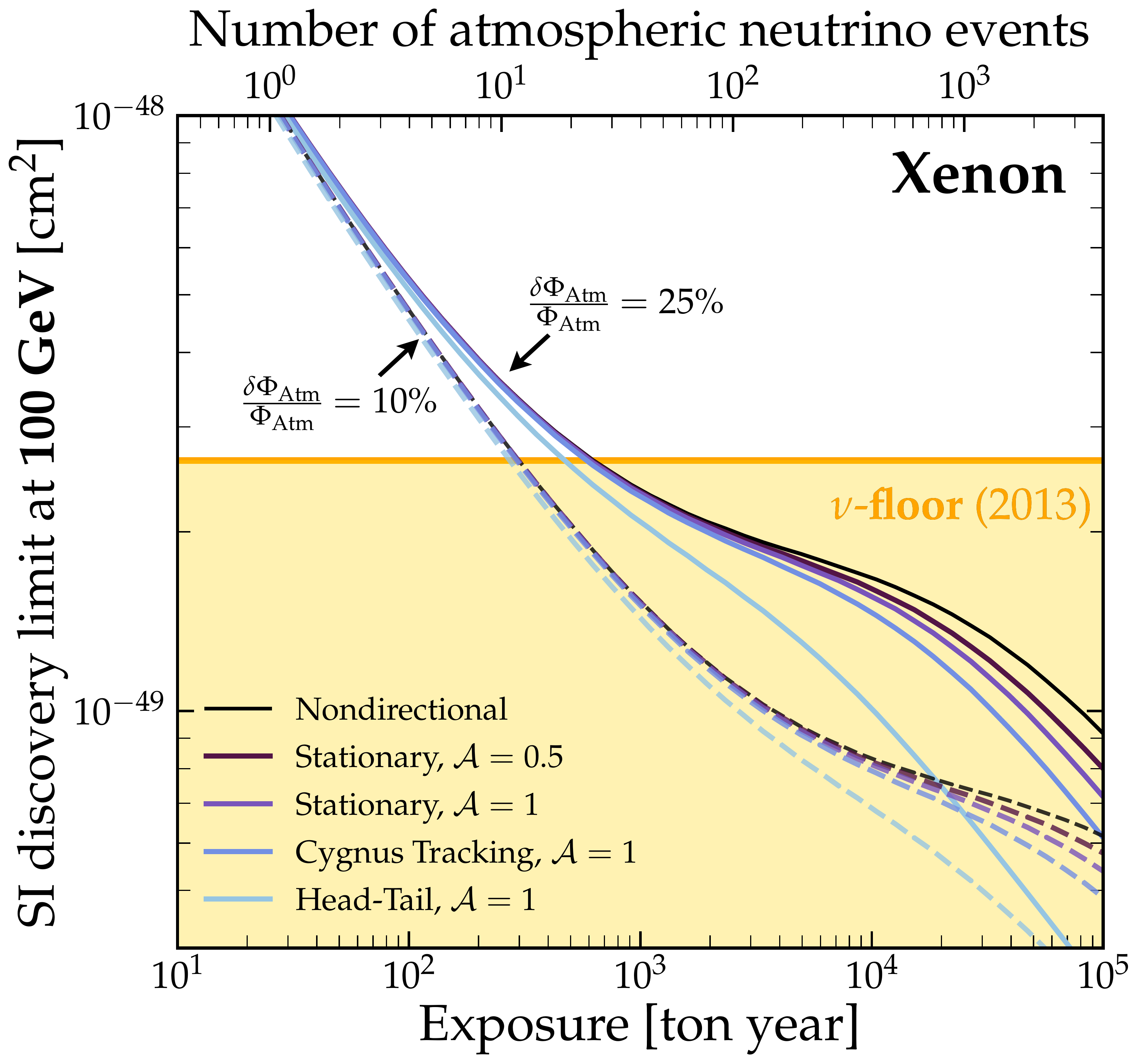}
\includegraphics[trim = 0mm 0 0mm 0mm, clip, width=0.49\textwidth]{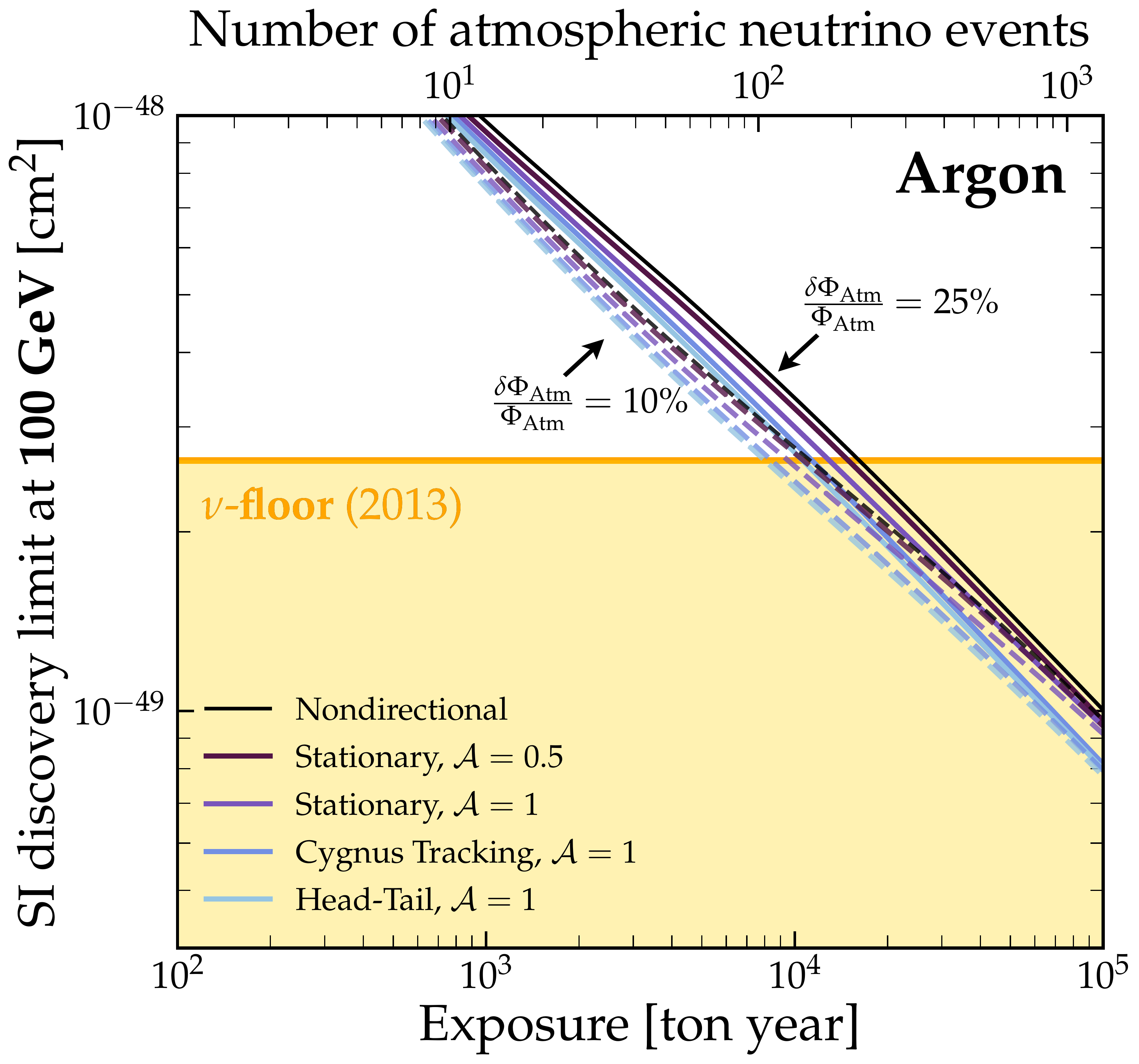}
\caption{SI discovery limits versus exposure at a fixed mass of 100 GeV, for xenon and argon target nuclei respectively. We show limits for five scenarios with increasing levels of directional information (black to light blue), explained in more detail in the text. For both panels we show two sets of lines corresponding to analyses assuming two different values for the atmospheric neutrino flux uncertainty: 25\% and 10\%, solid and dashed lines respectively.} 
\label{fig:discoverylimits}
\end{center}
\end{figure*} 

\begin{figure}
\begin{center}
\includegraphics[trim = 0mm 0 0mm 0mm, clip, width=0.49\textwidth]{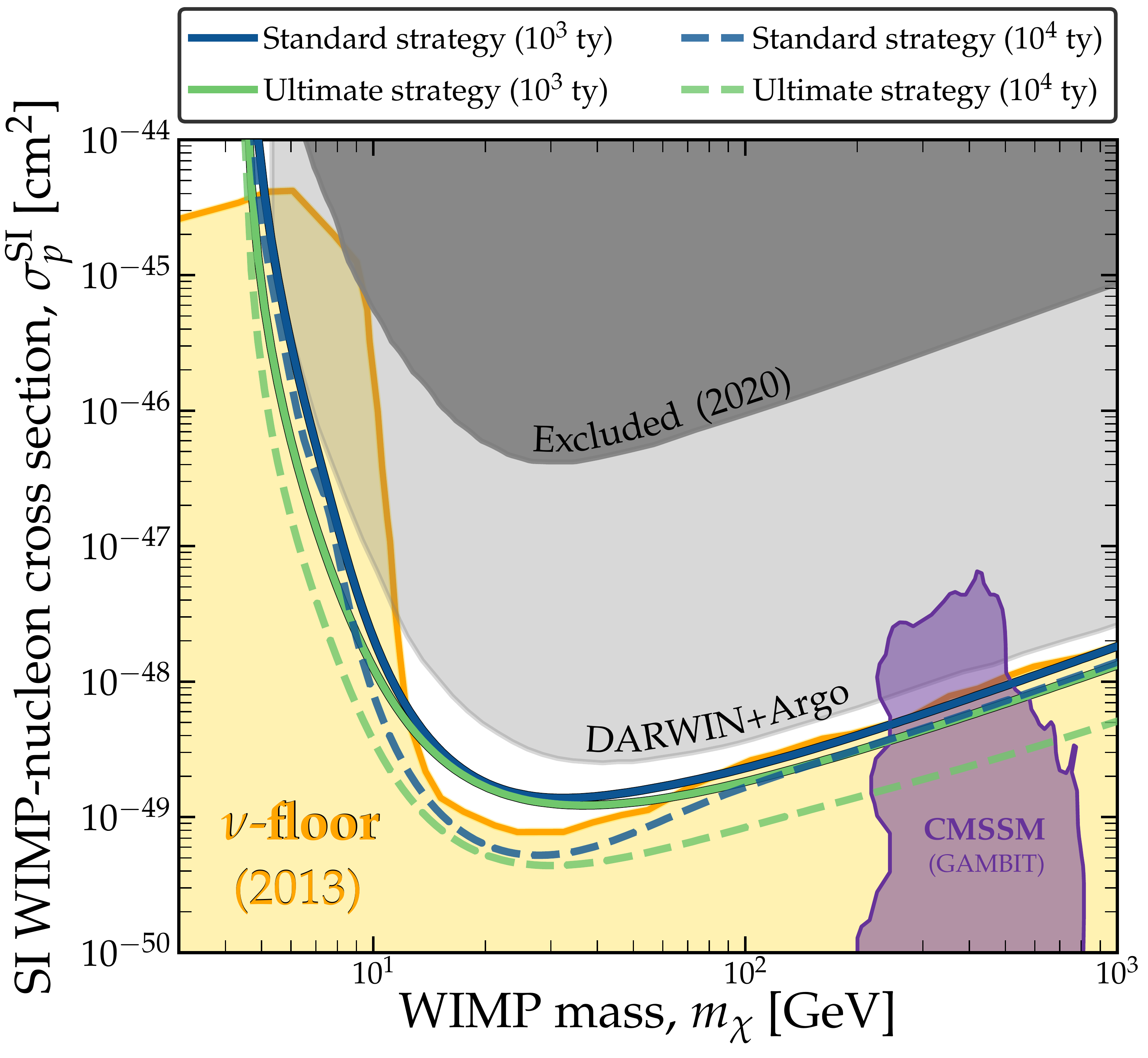}
\caption{SI discovery limits as a function of WIMP mass for two different strategies for probing at and below the neutrino floor. The ``standard strategy'' (blue lines) assume that the only possible discriminant between the WIMP signal and the neutrino background involves recoil energy. The ``ultimate strategy'' demonstrates the best potential discrimination between WIMPs and neutrinos in multiton-scale liquid noble experiments by exploiting all possible \emph{nondirectional} discriminants: improved atmospheric neutrino flux measurements from 25\% to 10\%; target complementarity with xenon and argon recoil information; and finally, the annual modulation signals. In the ultimate strategy, the exposure refers to the total combined xenon and argon exposure. As an example of WIMP candidates that could be discovered with this new strategy, we have taken a region of expected neutralino cross sections and masses from a GAMBIT global analysis of the CMSSM~\cite{Athron:2017qdc} (specifically, neutralinos from the scenario in which the relic density is not exceeded due to primarily stop coannihilations). We show the envelope of the projected sensitivities for DARWIN and Argo in light gray.} 
\label{fig:discoverylimits2}
\end{center}
\end{figure} 

\subsection{Columnar recombination vs the neutrino floor}
We will quantify how well columnar recombination allows the WIMP and atmospheric neutrino signals to be discriminated in both xenon and argon experiments (Fig.~\ref{fig:discoverylimits}), and then show how far below the atmospheric neutrino floor we could reasonably expect to be able to probe in the future, if we combine all non-directional discriminating information available (Fig.~\ref{fig:discoverylimits2}).

In Fig.~\ref{fig:discoverylimits} we show the evolution of the SI discovery limit for increasing exposures at a fixed mass of 100 GeV. To demonstrate how the discrimination power improves with increased directionality, we consider six scenarios:
\begin{itemize}
\setlength\itemsep{0.0em}
\item{{\bf Nondirectional} ($\mathcal{A} = 0$). No directional sensitivity: a reference point corresponding to a conventional LXe or LAr TPC.}
\item{{\bf Stationary} ($\mathcal{A} = 0.5$ and $1$). An experiment sensitive to columnar recombination, with asymmetry $\mathcal{A}$. The lower case $\mathcal{A}=0.5$ is approximately ten times stronger than the only existing experimental result at the relevant energy scales, shown at low significance by SCENE~\cite{Cao:2014gns}.}
\item{{\bf Cygnus-tracking} ($\mathcal{A}=1$). An experiment exploiting columnar recombination that also has some tracking mechanism to follow Cygnus across the sky, i.e. the electric field always aligns with the DM wind. This essentially makes maximal use of the direction dependence of columnar recombination.} 
\item{{\bf Head-Tail} ($\mathcal{A}=1$). Allowing for the discrimination of the sign of $\pm \cos{\theta}$. Columnar recombination has no Head-tail sensitivity; this scenario simply allows us to explicitly show how much sensitivity is lost because of this fact.} 
\end{itemize}  
We reiterate that our results correspond to an extrapolation beyond what could be realistically achieved currently. While there may turn out to be ways to improve upon existing results---such as the additions of TMA or TEA mentioned previously---this is an area which still requires a dedicated investigation.  Nevertheless our results are enough for now to help inform us as to what are the dominant limitations in preventing the technique from being more powerful.

As a concrete benchmark cross section, we compare our results with the typically quoted 2013 neutrino floor~\cite{Billard:2013qya}. From Fig.~\ref{fig:discoverylimits} we see that columnar recombination does help discriminate between the neutrino background as it should. But the degree to which it helps is somewhat disappointing. Overall, the improvement in discrimination power brought by directionality is relatively mild and begins to take effect only for quite large numbers of expected background events. For xenon this would require exposures up to $10^4$ ton-year and beyond, which is especially daunting given that columnar recombination may be unobservable in liquid xenon at the low energies needed here, and could require a high pressure gas mode. In argon the improvement is even more slight, though in this case the saturation of the WIMP signal by the neutrino background is much less severe, as discussed in Sec.~\ref{sec:neutrinofloor}. Comparing the ``Stationary'', and even the ``Cygnus-tracking'' limits with the "Head-Tail" case for xenon, we can can clearly see that it is the lack of recoil vector sense recognition that is the major limitation, as we anticipated earlier.

Also in Fig.~\ref{fig:discoverylimits} we repeat the analysis of each scenario to produce a second set of discovery limits (dashed lines) in which we assume the atmospheric flux uncertainty is reduced from 25\% to 10\%.  For argon this leads to only a mild improvement, but in xenon this complementary information could be of substantial benefit. In this case, an atmospheric flux measurement for neutrino energies below 100 MeV would be much more desirable than attempting to tease out a weak directional signature in a high-pressure gas mode.

\subsection{Ultimate reach of liquid noble experiments}

Since we have shown that columnar recombination may not be as powerful as previously suggested in probing beyond the neutrino floor. We will conclude by determining just how far a combination of xenon and argon experiments could probe. Finally we show how all the features of the WIMP and neutrino signals could be combined in an ultimate strategy to probe beyond the neutrino floor. We combine all possible nondirectional discriminants that may be accessible to LAr and LXe experiments in the future. These are:
\begin{itemize}
\setlength\itemsep{0.0em}
\item{{\bf Annual modulation:} exploiting the small annual modulations of the WIMP and solar neutrino event rates, as well as the small seasonal modulation of the angular dependence of the all-flavour atmospheric neutrino flux.}
\item{{\bf Target complementarity:} combining the slightly different recoil energy distributions from both xenon and argon experiments.}
\item{{\bf Improved atmospheric flux measurements:}  exploiting the complementarity with future neutrino telescopes. We assume a 10\% flux uncertainty as a potential improvement.}
\end{itemize}
In Fig.~\ref{fig:discoverylimits2} we compare the potential for this ``ultimate'' strategy with the ``standard'' strategy, i.e. using only recoil energy information from a xenon experiment. As in the previous figure we take the 2013 neutrino floor as a reference point. We compare two exposures of $10^3$ and $10^4$ ton-years which refers to the total combined exposure.

In Fig.~\ref{fig:discoverylimits2}, our standard strategy limits, shown in blue, are calculated in precisely the same way as this limit. Under the standard strategy an increase in exposure time by a factor of ten converts into a factor of only 1.3 improvement in the discovery limit for high masses. On the other hand, under the ultimate strategy the improvement is by a factor of 3.2, or in other terms a factor of 5.3 below the neutrino floor. While an exposure of $10^4$ is a stage beyond the next generation of multiton-scale detector, this result shows that under our ultimate strategy, the neutrino floor is decisively not the final limit to the direct detection of $\gtrsim 100$ GeV WIMP masses. 

These improvements could be especially valuable for discovering WIMP candidates that are still theoretically well-motivated, but are considered too difficult to reach. We show an example of where such a WIMP could lie, using a result from a recent GAMBIT~\cite{Athron:2017ard} global fit of GUT-scale SUSY~\cite{Athron:2017qdc}. We take the 2$\sigma$ confidence region for DM-neutralinos in the constrained minimal supersymmetric model (CMSSM). This particular scenario shown---in which the relic density is restrained by stop coannihilations---extends substantially below the neutrino floor. This is but one example of a model configuration that could greatly benefit from the strategy presented here. Several others exist in the same analysis of minimal SUSY, namely the non-universal Higgs mass models, and in other analyses of both SUSY~\cite{Roszkowski:2014iqa,Athron:2017qdc,Hisano:2011cs,Kobakhidze:2018vuy} and non-SUSY~\cite{Arcadi:2017wqi,Baker:2019ndr,Arina:2019tib} WIMPs.

\section{Discussion and Summary}\label{sec:summary}

We have asked whether it is possible to probe beyond the neutrino floor for masses $\gtrsim$100 GeV. There is substantial theoretical motivation for DM-nucleus cross sections in this mass range, and is a common region to find supersymmetric thermal WIMP candidates~\cite{Arcadi:2017kky,Roszkowski:2014iqa,Hisano:2011cs,Abdullah:2016avr,Athron:2017qdc,Kobakhidze:2018vuy}. While directional gas TPCs are potentially powerful for discriminating WIMPs and solar neutrinos~\cite{Grothaus:2014hja,O'Hare:2015mda}, relevant at lower masses, they would require infeasibly large volumes to reach the neutrino floor at higher masses.

Here we have instead looked towards maximising the discovery potential of multiton-scale xenon and argon-based experiments. Our final result Fig.~\ref{fig:discoverylimits2} shows how all possible discriminants can be be combined to discriminate against the atmospheric neutrino background: annual modulation, target complementarity, improved flux measurements, and directionality. The latter discriminant in particular has been the focus of this work and is inspired by past and ongoing investigations to extract a directional effect in xenon and argon: columnar recombination~\cite{Nygren:2013nda}. At this stage, the effect still requires further experimental verification~\cite{Cao:2014gns} but we have implemented a simplified model to be able to quantify how much it could help discriminate against the atmospheric neutrino background. Because we have made several idealised assumptions our results should be considered optimistic. However despite this, we can conclude that directionality may in fact not be the most viable strategy for probing beyond the high mass neutrino floor, in contrast to some previous claims as to the potential utility of columnar recombination~\cite{Cadeddu:2017ebu}. 
Ultimately the columnar recombination signal is most greatly harmed by its lack of head-tail sensitivity.

Putting potential directional signals aside, we have also shown here that the neutrino floor at high masses need not be the final limit to liquid noble experiments. Interestingly, we found that any slight improvement in sensitivity brought by directionality would be far outweighed by an improved measurement of the atmospheric flux to a level of 10\%. We have also shown that a joint analysis, combining 100--1000 ton-year xenon and argon experiments in the near future, could push the sensitivity to ultra-low WIMP cross sections even further below the neutrino floor. This result clearly shows that the experimental efforts towards developing both argon and xenon TPCs are both highly worthwhile. Potentially, one could imagine that the final stage of WIMP direct detection might involve a global coordination of all the large-scale liquid noble experiments. Since the neutrino floor at high masses is surmountable, those well-motivated theoretical predictions for WIMP cross sections may turn out to be within reach after all.

So are atmospheric neutrino flux measurements at the 10\% level assumed in Figs.~\ref{fig:discoverylimits} and~\ref{fig:discoverylimits2} foreseeable? This would require that at least $\mathcal{O}(100)$ atmospheric neutrinos could be detected in the troublesome regime below 100 MeV, this would be achievable in experiments such as DUNE~\cite{Abi:2020evt} and JUNO~\cite{An:2015jdp}. Taking the case of DUNE~\cite{Abi:2020evt}, since it is a LAr TPC it is expected to have particularly good sensitivity at low energies to the $\nu_e$ component of the flux via the charged current (CC) interaction $\nu_e + ^{40}$Ar$ \rightarrow {}^{40}{\rm K}^* + e^{-}$. DUNE plans to have a mass of 20 kiloton in operation by 2024, and a common final benchmark exposure is 350 kiloton-years (two 14,490 m$^3$ modules operating for ten years). A precise calculation of the projected atmospheric neutrino flux measurement would be background dependent~\cite{Cocco:2004ac,Zhu:2018rwc}, and beyond the scope of this work. However we can make a simple calculation of the atmospheric neutrino event rate for DUNE by combining the $\nu_e$-$^{40}$Ar CC cross section; the FLUKA atmospheric $\nu_e$ flux below 100 MeV; and DUNE's sensitivity for CC events which sets a threshold of $\sim$9 MeV (see e.g. Ref.~\cite{Capozzi:2018dat}). We arrive at $\sim$330 events in 350 kiloton-years, though there will be more from other interactions and from $\nu_\mu$. So an uncertainty between 25\%--10\% for the all-flavour flux would certainly be realistic within a 10 year exposure, though this is still very low compared with the rate of higher energy neutrinos.  There are also degeneracies between neutrino oscillation parameters, so a detailed projection for this would require a combined analysis with other experiments, see e.g.~\cite{Dutta:2020che}. In particular we should also anticipate atmospheric neutrino events at the very low end of the energy sensitivity projected for JUNO~\cite{An:2015jdp}, though smaller in number than in DUNE.

Ultimately, it seems realistic to anticipate a reduction in the neutrino flux uncertainties. This could pave the way for the neutrino floor at high masses to eventually be overcome with a coordination of massive liquid xenon and argon experiments detecting both neutrinos and dark matter.


\acknowledgments
 C.A.J.O. is supported by the University of Sydney and the Australian Research Council.

\maketitle
\flushbottom

\bibliographystyle{bibi}
\bibliography{HighMassNuFloor}

\end{document}